\documentclass[twocolumngrid,prd,superscriptsize,reprint]{revtex4-1}
\usepackage[utf8]{inputenc} 
\usepackage{graphicx,xcolor,overpic,mathtools}
\usepackage{amsthm,amsmath,amssymb}
\usepackage[colorlinks]{hyperref}
\usepackage{textcomp}
\definecolor{coolblack}{rgb}{0.0, 0.18, 0.39}
\hypersetup{
    colorlinks = false,
   }
\PassOptionsToPackage{normalem}{ulem}
\usepackage{ulem} 
\usepackage{sidenotes}
\usepackage{bm}
\usepackage{url}
\usepackage{todonotes}
\usepackage{physics}
\usepackage[makeroom]{cancel}
\usepackage{cleveref}
\usepackage{tensor}
\usepackage{mathrsfs}
\usepackage{slashed}
\usepackage{caption}
\usepackage{subcaption}
\usepackage[mode=buildnew]{standalone}

\newcommand{\comment}[1]{}

\usepackage{pgfplots}

\usepackage{diagbox}
\usepackage{stackengine,amssymb,graphicx}

\NewDocumentCommand{\evat}{sO{\bigg}mm}{%
  \IfBooleanTF{#1}
   {\mleft. #3 \mright|_{#4}}
   {#3#2|_{#4}}%
}
\usepackage{enumerate}
\definecolor{azure}{rgb}{0.0, 0.5, 1.0}

\begin{document}

\title[]{Galactic Halos and rotating bosonic dark matter}

\author{Jorge Castelo Mourelle}
\author{Christoph Adam}

\affiliation{%
Departamento de F\'isica de Part\'iculas, Universidad de Santiago de Compostela and Instituto
Galego de F\'isica de Altas Enerxias (IGFAE) E-15782 Santiago de Compostela, Spain
}%

\date[ Date: ]{\today}
\begin{abstract}

Rotating bosonic dark matter halos are considered as potential candidates for modeling dark matter in galactic halos. These bosonic dark matter halos can be viewed as a dilute and very extended version of bosonic stars,
and the methods used for the calculation and analysis of the latter objects can be directly applied.
Bosonic
stars, a hypothetical type of astrophysical objects, are categorized into two primary families, based on the nature of the particles composing them: Einstein-Klein-Gordon stars and Proca stars. We examine various models from both families and the rotation curves which their contribution induces in different galaxies, to identify the most plausible candidates that explain the flattening of orbital velocities observed in galactic halos. By exploring different combinations of our dark matter models with observable galactic features, we propose an interesting source to compensate for the apparent lack of matter in dwarf and spiral galaxies, providing a possible explanation for this longstanding astronomical puzzle.

\end{abstract}

\maketitle

\begin{quote}
 
\end{quote}

%\begin{minipage}{\textwidth}
%\tableofcontents
%\end{minipage}

%%%%%%%%%%%%%%%%%%%%%%%%%%%%%%%%%%%%%%%%%%%%%%%%%%%%%%%%%%%%%%
\section{Introduction}
%\paragraph*{Introduction.---}
%%%%%%%%%%%%%%%%%%%%%%%%%%%%%%%%%%%%%%%%%%%%%%%%%%%%%%%%%%%%%%

According to theoretical predictions, rotation curves of galaxies and galaxy clusters should exhibit a Keplerian decline, following the formula $v \sim \sqrt{1/r}$ between the velocity $v$ and the distance $r$ from the center outside the zones occupied by observable, luminous matter. However, contrary to these predictions, the rotation curves measured in these regions do not taper off but instead remain unexpectedly flat, extending far beyond the galaxies' most distant visible sectors \cite{peebels}. This anomaly, often referred to as the Rotation Curves (RC) \cite{doi:10.1126/science.220.4604.1339,Persic:1995ru,Primack:2001ia} problem, is a key piece of evidence supporting the existence of non-baryonic dark matter.

Various theories have been proposed to explain these anomalies in galaxy rotation curves. Some suggest modifying Newtonian gravity \cite{1983ApJ...270..365M}, while others support the existence of invisible, non-interacting dark matter \cite{Chardin:1999jj,Baltz:1997dw,Beylin:2020bsz}. Recent studies across diverse gravitational models have underscored the critical role of dark matter in accounting for these phenomena \cite{Zhytnikov:1994rm}. Moreover, the notion that massive compact halo objects (MACHOs), composed of baryonic matter, could explain these anomalies is increasingly being dismissed \cite{Weinberg:1996qh,Bai:2020jfm,Griest:1993cv}.

Fritz Zwicky first proposed the existence of dark matter in galaxy clusters, observing that the Coma cluster's gravitational mass far exceeded that of its luminous matter, by roughly an order of magnitude \cite{2009GReGr..41..207Z}. By the early 1970s, enhanced measurements of galaxy rotation curves indicated high mass-to-luminosity ratios, especially beyond certain radii, pointing to significant amounts of invisible matter \cite{1970ApJ...159..379R}. Freeman's theory of spherical halos around galaxies introduced the concept of a linearly increasing mass function within these structures \cite{1970ApJ...160..811F}. Subsequent research, leveraging techniques like satellite galaxy motion analysis, weak gravitational lensing, and quasar absorption lines, has further probed the extensions of dark matter halos, sometimes finding radii exceeding $400$ $\mathrm{kpc}$. \cite{Persic:1995ru,Brainerd:1995da,Lanzetta_1996,Primack:1997av,Zaritsky:1996ch}.

Bosonic fields coupled to gravity are categorized primarily into Boson Stars (BS) and Proca Stars (PS), modeled as massive free or self-interacting complex scalar and complex vectorial fields hold together by gravity, respectively \cite{PhysRev.172.1331,PhysRev.187.1767,PhysRev.148.1269,Schunck:2003kk,Brito:2015pxa,Herdeiro:2023wqf}. Recently, research has expanded to Proca-Higgs stars (PHS), which combine complex vectors with real scalars for enhanced field interactions \cite{Herdeiro:2023lze,Brito:2024biy}. Additionally, the exploration of multi-field configurations has led to the development of multi-state boson stars and $\ell$-Bosonic stars, further diversifying potential models \cite{Urena-Lopez:2010zva,Sanchis-Gual:2021edp,Alcubierre:2018ahf,Lazarte:2024jyr}.

Originating from concepts like geons and spherical BS, bosonic stars have been extensively studied for their properties, stability, and formation \cite{PhysRev.97.511,PhysRev.172.1331,PhysRev.187.1767,PhysRev.148.1269,Liebling:2012fv,Lai:2004fw,Schunck:2003kk}. Traditionally envisioned as static and spherically symmetric, bosonic fields coupled to gravity are now widely modeled as rotating, forming axisymmetric spinning configurations in both scalar and vector forms \cite{Schunck:1996he,Yoshida:1997qf,Brito:2015pxa,Herdeiro:2016tmi,Herdeiro:2019mbz}. Extensive research into these systems has not only explored their physical properties and stability but also their dynamic behaviors, significantly enhancing our understanding within astrophysics \cite{Vincent:2015xta,Delgado:2021jxd,Delgado:2022pwo,Sengo:2024pwk,Sanchis-Gual:2019ljs,Sanchis-Gual:2021phr,Siemonsen:2020hcg}. Their versatility, shaped by the choice of matter fields and Lagrangian potentials, allows them to mimic a variety of astrophysical objects such as Neutron Stars (NS), Black Holes and intermediate-mass objects making them an invaluable tool in astrophysics \cite{Schunck:1998nq,PhysRevD.80.084023,Herdeiro:2021lwl,Rosa:2022tfv,Rosa:2023qcv,Sengo:2024pwk,Adam:2010rrj}. 

Several studies have explored the potential of static scalar bosonic matter coupled to gravity as a dark matter source \cite{Schunck:1998nq,Urena-Lopez:2010zva,Broadhurst:2019fsl,Chen:2020cef,Annulli:2020ilw,Amruth:2023xqj,Pozo:2023zmx}. Models in which an ultralight axion-like particle is responsible for the dark matter in the galaxy take on special relevance \cite{Luu:2018afg,deMartino:2018krg,Emami:2018rxq,2024arXiv240104735K,Sakharov:2021dim}. The Scalar Field Dark Matter model, aligned with the $\Lambda$CDM model, predicts large-scale phenomena consistent up to linear order of perturbations \cite{PhysRevD.68.024023,PhysRevD.35.3640,Sahni:1999qe,Matos:2000ss,Matos:2008ag,Urena-Lopez:2010zva}. This model has also been used to examine issues related to rotation curves through the concept of scalar fields \cite{Sin:1992bg,Ji:1994xh,PhysRevD.69.124033,Lee:1995af}, describing dark matter halos as Newtonian Bose-Einstein Condensates. The modeling of these halos uses the ground state solutions of the Schrödinger-Poisson (SP) system, the only stable configuration where all boson particles are in the ground state. These configurations marginally align with observed rotation curves as supported by multiple studies \cite{Ji:1994xh,PhysRevD.69.124033,Arbey:2003sj,Guzman:2006yc}. Stability analyses further confirm that the ground state is resilient to gravitational perturbations, whereas excited state configurations are inherently unstable \cite{Ruffini:1969qy,Guzman:2004wj,Schunck:2003kk}.

This paper investigates the possibilities of modelizing DM halos with different {\em rotating} bosonic fields. 
Galaxies and other structures which are formed by accretion or collapse of some kind of matter under its own gravitational attraction will generically end up in configurations with a non-zero angular momentum, i.e., they will form rotating objects. The case of rotating (dark) matter, therefore, is the relevant one in general, whereas the static, non-rotating case corresponds to a limit. 

Further, a stable configuration of rotating bosonic matter should lead to a stationary and axially symmetric space-time metric, and this condition restricts the possible dependence of the bosonic field on time $t$ and on the azimutal angle $\psi$ to $\Phi \sim e^{i(n\psi + \omega t)}$, where $\omega $ is the internal field frequency and $n$ is the integer azimutal winding number. 
As a consequence of the integer-valuedness of $n$, the transition from static ($n=0$) to rotating ($n\not= 0$) configurations is non-trivial and is possible only under certain averaging assumptions over the angular components of the system \cite{Kobayashi:1994qi,Delgado:2024fhc}. This means that traditional formalisms like Hartle-Thorne or slow-rotation approaches \cite{Hartle:1967he,Hartle:1968si} are not applicable, in general, and the rotating solutions differ significantly from the static ones.

We employ various systems of scalar and vectorial bosonic fields coupled to gravity in an axisymmetric framework to look for good candidates when modeling the non-luminous part of various rotation curves. Despite using full General Relativity calculations, we shall find that all relevant cases belong to the non-relativistic regime.

The structure of the paper is as follows: In  \cref{setup} we introduce
the theoretical set-up for the models under study. We present the numerical scheme in  \cref{numerical} and introduce the system of units in \cref{rescalingsect}. In \cref{geodesicssect}, we present the different orbits, which allow us to obtain the rotation curves.  In \cref{models}, we show how to obtain the different contributions to the RC through the luminous and DM models. In \cref{results}, we present our modelization together with the observational data with the aim of a comparison. Section \ref{conclusions} is devoted to conclusions, and we also show a set of useful equations for understanding the method of solving and some properties in \cref{appendixA}.

%%%%%%%%%%%%%%%%%%%%%%%%%%%%%%%%%%%%%%%%%%%%%%%%%%%%%%%%%%%%%%

%%%%%%%%%%%%%%%%%%%%%%%%%%%%%%%%%%%%%%%%%%%%%%%%%%%%%%%%%%%%%%
\section{Spinning bosonic fields coupled to gravity}\label{setup}
%%%%%%%%%%%%%%%%%%%%%%%%%%%%%%%%%%%%%%%%%%%%%%%%%%%%%%%%%%%%%%

In this approach, we describe scalar and vectorial boson stars using the Einstein-Klein-Gordon (EKG) and Einstein-Proca (EP) frameworks, where a massive complex scalar field $\Phi$ or a vectorial field $A$ is minimally coupled to Einstein's gravity, as outlined in the following action \cite{Liebling:2012fv}:
\begin{equation}
\mathcal{S}_{(\Phi,A)} = \int \left(\frac{1}{16\pi G} R + \mathcal{L}_{(\Phi,A)} \right) \sqrt{-g} d^4x.
\label{action}
\end{equation}
Here,
$g$ represents the determinant of the metric,
$R$ is the Ricci scalar, and $\mathcal{L}_{(\Phi,A)}$ are the Lagrangians governing the dynamics of the scalar and vectorial fields $\Phi$ and $A$.

The systems under study should consider rotation, and for this reason, we assume the following ansatz for the metric, describing the stationary and axisymmetric space-time \cite{Herdeiro:2015gia,PhysRevD.55.6081}:
\begin{equation}
\begin{split}
ds^2=&-e^{2\nu}dt^2+e^{2\beta}r^2\sin^2\theta\left(d\psi-\frac{W}{r}dt\right)^2\\
&+e^{2\alpha}(dr^2+r^2d\theta^2),
\label{axmetric}
\end{split}
\end{equation}
where $\nu$, $\alpha$, $\beta$, and $W$ are functions dependent only on $r$ and $\theta$.

%%%%%%%%%%%%%%%%%%%%%%%%%%%%%%%%%%%%%%%%%%%%%%%%%%%%%%%%%%%%%%
\subsection{Scalar Boson Stars}
%%%%%%%%%%%%%%%%%%%%%%%%%%%%%%%%%%%%%%%%%%%%%%%%%%%%%%%%%%%%%%
The Lagrangian density describing the EKG system is as follows:

\begin{equation}
\mathcal{L}_{\Phi} = -\frac{1}{2} \left[g^{\alpha\beta} \nabla_{\alpha} \Phi^* \nabla_{\beta} \Phi + V\left(|\Phi|^2\right) \right].
\label{lagrangianscalar}
\end{equation}

This model maintains a global $U(1)$ symmetry, meaning that the potential $V\left(|\Phi|^2\right)$ depends solely on the modulus of the scalar field. Each potential under consideration incorporates the quadratic mass term $\mu^2|\Phi|^2$ along with additional interaction terms, allowing for various BS configurations through different self-interacting terms in the potential. Analogous to the equation of state (EOS) in NS studies, the scalar potential for BS determines the characteristics of these stars. For this analysis, we employ the self-interacting potential models presented in \cref{curvesmodelsbs}.

Upon varying the action \cref{action} with the lagrangian density \cref{lagrangianscalar}, the EKG equations are derived,
\begin{equation}
\begin{split}
    &R_{\alpha\beta}-\frac{1}{2}Rg_{\alpha\beta}=8\pi T^{\Phi}_{\alpha\beta}, \\
    &
    g^{\alpha\beta}\nabla_\alpha\nabla_{\beta}\Phi=\frac{dV}{d|\Phi|^2}\Phi,
\label{kg}
\end{split}
\end{equation}
where 
$R_{\alpha\beta}$ denotes the Ricci tensor and $T^{\Phi}_{\alpha\beta}$ is the canonical Stress-Energy tensor for the scalar field, which is defined as
\begin{equation}
\begin{split}
T_{\alpha\beta}^{\Phi}=&\nabla_{\alpha}\Phi^*\nabla_{\beta}\Phi+\nabla_{\beta}\Phi^*\nabla_{\alpha}\Phi-\\
&g_{\alpha\beta}\left[g^{\mu\nu}\left(\nabla_{\mu}\Phi^*\nabla_{\nu}\Phi+\nabla_{\nu}\Phi^*\nabla_{\mu}\Phi\right)+V\left(|\Phi|^2\right)\right].
    \label{stress}
    \end{split}
\end{equation}
To ensure the stationarity and axial symmetry required by the Stress-Energy tensor, the ansatz represents the scalar field:
\begin{equation}
     \Phi(t,r,\theta,\psi)=\phi(r,\theta)e^{-i(w t+n\psi)}.
     \label{scalar}
 \end{equation}
Here $w \in \mathbb{R}$ is the angular frequency of the field, and $n \in \mathbb{Z}$ (also called $m$ or  $s$ in the literature \cite{Vaglio:2022flq,Ryan:1996nk}) is the \textit{azimutal harmonic index}, also called \textit{azimutal winding number}. This parameter enters the problem as an integer related to the star's angular momentum and is required to describe rotating BS. Further, $\phi(r,\theta)$ is the profile of the star.

%%%%%%%%%%%%%%%%%%%%%%%%%%%%%%%%%%%%%%%
\subsection{Vectorial  Boson Stars}
%%%%%%%%%%%%%%%%%%%%%%%%%%%%%%%%%%%%%%%

The Lagrangian density which describes the EP system is the following \cite{Herdeiro:2016tmi,Brito:2015pxa},

\begin{equation}
    \mathcal{L}_A=-\frac{1}{4}\textit{F}_{\alpha\beta}\bar{\textit{F}}^{\alpha\beta}-\frac{1}{2}\mu^2\textit{A}_{\alpha}\bar{\textit{A}}^{\alpha},
\end{equation}
where $F_{\alpha\beta}=\partial_{\alpha}A_{\beta}-\partial_{\beta}A_{\alpha}$. Further, the Euler-Lagrange equations imply the Lorenz condition $\nabla_{\alpha}\textit{A}^{\alpha}=0$. We will also occasionaly use the one-form $A= A_\mu dx^\mu$ associated with the covector $A_\mu$.

The Einstein-Proca system is described by the equations
\begin{equation}
\begin{split}
    &R_{\alpha\beta}-\frac{1}{2}Rg_{\alpha\beta}=8\pi T^{A}_{\alpha\beta}, \\
    &
    \nabla_{\alpha}\textit{F}^{\alpha\beta}=\mu^2\textit{A}^{\beta},
\label{pr}
\end{split}
\end{equation}

%We use the same metric ansatz \cref{metric} than for the scalar case.
The equations of motion are
\begin{equation}
\textit{F}^{\alpha\beta}_{,\alpha}+\Gamma^{\alpha}_{\mu\alpha}\textit{F}^{\mu\beta}+\Gamma^{\beta}_{\mu\alpha}\textit{F}^{\alpha\mu}-\mu^2A^{\beta}=0,
\end{equation}
and the Lorenz condition reads
\begin{equation}
    \mathcal{L}=A^{\alpha}_{,\alpha}+\Gamma^{\alpha}_{\mu\alpha}A^{\mu}=0.
    \label{lorenz}
\end{equation}

The axisymmetric spacetime described by \cref{axmetric} is compatible with the following Stress-Energy tensor for the vectorial case,
\begin{equation}
\begin{split}
    T^{A}_{\alpha\beta}=&-F_{\sigma(\alpha}\bar{F}_{\beta)}^{\sigma}-\frac{1}{4}g_{\alpha\beta}F_{\mu\nu}\bar{F}^{\mu\nu}+\\
    &
    \mu^2\left[A_{(\alpha}\bar{A}_{\beta)}-\frac{1}{2}g_{\alpha\beta}A_{\sigma}\bar{A}^{\sigma}\right],
\end{split}
\end{equation}

and with the field ansatz
\begin{equation}\label{ansatzProca}
    \textit{A}=e^{i(n\psi-w t)}\left(i V dt+\frac{H_1}{r}dr+H_2 d\theta+iH_3\sin\theta d\psi \right) ,
\end{equation}
where $V,H_1,H_2$ and $H_3$ depend on $r$ and
$\theta$ coordinates. $n$ is, again, the azimutal harmonic index. 
Here, we restrict to the case $n=1$, because it turns out that rotating EP halos for $n>1$ lead to rotation curves with velocities much higher than observed.

%%%%%%%%%%%%%%%%%%%%%%%%%%%%%%%%%%%%%%%%%%%%%%%%%%%%%%%%%%%%%%
\section{Numerical implementation for the EKG and EP models }\label{numerical}
%%%%%%%%%%%%%%%%%%%%%%%%%%%%%%%%%%%%%%%%%%%%%%%%%%%%%%%%%%%%%%

The numerical integration of the EKG and EP systems is done using normalization of the radial distance and angular frequency by the mass $\mu$ of the bosonic field, using the transformations $r\rightarrow r\mu$ and $w\rightarrow w/\mu$. The dependence on $\mu$ for the field equations is eliminated, obtaining numerical solutions in units of $\mu$. A specific physical value for the bosonic mass is recovered when we return to dimensional units for the observables, and the dimensional rescaling is explained in \cref{rescalingsect}.

Our two distinct problems are characterized by their own coupled, non-linear, partial differential equations. The scalar case involves five equations for the metric functions and the scalar field, while the vectorial case comprises eight equations, four for the metric functions and four for the field functions. We employ the FIDISOL/CADSOL package, a Newton-Raphson-based solver with a flexible grid and order of accuracy, providing an error estimate for each unknown function.

To handle the infinite range of the radial coordinate, we compactify it with $x \equiv r/(c+r)$, mapping $r$ from $[0,\infty)$ to a finite interval $x \in [0,1]$. In the BS models, we discretize the equations on a $(401 \times 40)$ $(x, \theta)$ grid, where $x$ ranges from 0 to 1 and $\theta$ from 0 to $\pi/2$. Here, we set $c=1$ due to the extensive grid size. For the Proca case, which requires more computational resources, we reduce the grid size but adjust the $c$ value to ensure sufficient resolution in the far-field region, depending on the frequency ranges where solutions are sought.

\subsection{Boundary conditions}
Asymptotic flatness for the field profile and the metric functions leads to, 
\begin{equation}
    \lim_{r\rightarrow\infty} \alpha =\lim_{r\rightarrow\infty} \beta =\lim_{r\rightarrow\infty} \nu =\lim_{r\rightarrow\infty} W =\lim_{r\rightarrow\infty} \phi = 0.
\end{equation}
Reflection symmetry on the rotation axis and  axial symmetry at $\theta=0$ and $\theta=\pi$ imply
\begin{eqnarray}
    \partial_{\theta}\alpha=\partial_{\theta}\beta=\partial_{\theta}\nu=\partial_{\theta}W=\partial_{\theta}\phi=0. 
\end{eqnarray}

The previous conditions must also hold on the equatorial plane $\theta=\pi/2$ due to the symmetry with respect to a reflection along the equator.
Regularity at the origin requires $\partial_r \alpha=\partial_r \beta=\partial_r \nu =W= \phi=0$ when $r \to 0$, and regularity in the symmetry axis further imposes $\left.  \alpha=\beta \right|_{\theta=0, \pi}$ \cite{Herdeiro:2015gia}.
Further details about the solver are explained in  \cite{Delgado:2022pwo,Adam:2022nlq}.

In the vectorial case, we have to impose, at the origin, 
\begin{equation}
\begin{split}
&\partial_r\alpha|_{r=0}=\partial_r\beta|_{r=0}=\partial_r\nu|_{r=0}= \\&
    W|_{r=0}=H_i|_{r=0}=V|_{r=0}=0, 
   \end{split}
\end{equation}
and at infinity 
\begin{equation}
 \begin{split}
   &\lim_{r\rightarrow\infty} \alpha=\lim_{r\rightarrow\infty} \beta=\lim_{r\rightarrow\infty} \nu=\\  &\lim_{r\rightarrow\infty}W=\lim_{r\rightarrow\infty}H_i=\lim_{r\rightarrow\infty}V=0.
   \end{split}
 \end{equation}
On the symmetry axis, the boundary conditions are
\begin{equation}
\begin{split}
&\partial_{\theta}\alpha|_{\theta=0,\pi}=\partial_{\theta}\beta|_{\theta=0,\pi}=\partial_{\theta}\nu|_{\theta=0,\pi}=\\
&\partial_{\theta}W|_{\theta=0,\pi}=\partial_{\theta}H_{2,3}|_{\theta=0,\pi}=\\&H_1|_{\theta=0,\pi}=V|_{\theta=0,\pi}=0. 
\end{split}
\end{equation}

%%%%%%%%%%%%%%%%%%%%%%%%%%%%%%%%%%%%%%%%%%%%%%%%%%%%%%%%%%%%
\subsection{Physical units}
\label{rescalingsect}
%%%%%%%%%%%%%%%%%%%%%%%%%%%%%%%%%%%%%%%%%%%%%%%%%%%%%%%%%%%%
%The natural system of units $G = c = {\hbar} = 1$ is used in most of the paper. 
%Further,
We use dimensionless units and quantities rescaled by the boson mass
for our numerical calculations, which must be converted back to dimensionful expressions when we use our models to fit real rotation curve data. Further, we assume $ c = {\hbar} = 1$ in the paper unless stated otherwise. In the numerical calculations we also set $G=1$, which is equivalent to expressing masses, lengths and time in Planck units.
For convenience, we briefly review the conversion factors from our numerical variables to the physical observables in the relevant astrophysical units.  The dimensionless variables, characterized by a tilde, are
\begin{equation} 
\Tilde{r} = r\mu, \quad \Tilde{t} = t\mu, 
%\quad \Tilde{M} = \frac{M}{\mu}, 
\quad \Tilde{\omega} = \frac{\omega}{\mu}. 
\end{equation}
The boson mass parameter $\mu$, as it appears in the bosonic lagrangians, naturally has the dimension of an inverse length and corresponds to the inverse of the reduced Compton wavelength of the boson, related to its mass $m$ by $m= (\hbar \mu)/c$. In practice, $m$ is frequently given in energy units (eV), and the conversion factor is
$1 \text{eV}/c^2 = 1.783 \times 10^{-36} \text{kg}$. Further, the conversion to the Compton wavelength in meters requires the values $\hbar = 1.054 \times 10^{-34} \text{ Js}$ and $c = 2.998 \times 10^8 \text{ m/s}$.
Astrophysical distances are usually provided in kiloparsecs, with the conversion factor
$1 \text{ kpc} = 3.086 \times 10^{19} \text{ m}$.

Finally, we need the behaviour of the full boson "star" (bosonic halo) mass $M$ under the rescaling by the boson mass $m$ (or $\mu$). One might naively expect that it scales inversely to the radial coordinate, but it scales, in fact, exactly like the radial coordinate, as a consequence of the Einstein equations. A simple way of understanding this may be achieved by a restriction to a static, spherically symmetric mass distribution. In Schwarzschild type coordinates, the corresponding metric can be expressed like $ds^2 = A(r) dt^2 + B(r) dr^2 + r^2 (d\theta^2 + \sin^2 \theta d\varphi^2 )$. Further, the metric function $B(r)$ can be expressed in terms of the total mass function $M(r)$ (the total mass enclosed within the radius $r$) via $B(r) = \left( 1- (2GM(r))/(c^2 r) \right)^{-1}$, and the scaling behaviour of $M(r)$ now follows from the scale invariance of $B(r)$.

That is to say, the total mass $M$ rescales like $\Tilde{M} = M\mu$ in Planck units, which in general units turns into (here $M_P$ is the Planck mass and $L_P$ the Planck length)
\begin{equation} 
\Tilde{M } = \frac{M\mu L_P}{M_P} = \frac{Mm}{M_P^2} \; , \quad M_P = \sqrt{\frac{\hbar c}{G}} = 2.1764 \cdot 10^{-8} \, \text{kg}.
\end{equation}
A widely used mass unit in astrophysics is the solar mass $M_\odot$, with the conversion factor $1M_\odot = 1.9884 \cdot 10^{30} \, \text{kg}$.

For the rotation curves, we also have to obtain the velocities in the desired units. Our dimensionless numerical calculations, obviously, provide the velocities in units of $c$, so we just have to multiply by $c= 299800 \mathrm{km/s}$ to express velocities in the astrophysical units of $\mathrm{km/s}$.

%%%%%%%%%%%%%%%%%%%%%%%%%%%%%%%%%%%%%%%%%%%%%%%%%%%%%%%%%%%%%%%
\section{Circular geodesics around bosonic stars}
\label{geodesicssect}
%%%%%%%%%%%%%%%%%%%%%%%%%%%%%%%%%%%%%%%%%%%%%%%%%%%%%%%%%%%%%%%
The full derivation of the orbit of a particle with arbitrary mass orbiting around an axisymmetric space-time, following \cite{Gourgoulhon:2010ju,Meliani:2015zta}, is shown in \Cref{appendixA}. This calculation is very important to determine the accretion dynamics for astrophysical relativistic objects. Already the existence of orbits is a non-trivial problem. For some regions, depending on the gravitational source, stable orbits are forbidden inside the delimitating innermost stable circular orbit (ISCO) radius $r_{ISCO}$, such that circular orbits only exist if $r>r_{ISCO}$. The velocity \cref{velocitystars} equation better explains this limit value \cite{Grandclement:2014msa}.

%% aren't they called ISCO instead of ICO?

The velocity for the matter orbiting equatorially our axisymmetric gravitational sources, as a function of the radius, is given by 

\begin{equation}
\begin{split}
    &v(r)\pm=\frac{e^{\beta-\nu }r\frac{\partial}{\partial r}\left(\frac{W}{r}\right)\pm\sqrt{D}}{2\left(\frac{\partial \beta}{\partial r}+\frac{1}{r}\right)},\\
    \end{split}
    \label{velocityhalo}
\end{equation}

with 
\begin{equation}
\begin{split}
   D=e^{2\beta-2\nu }r^2\left(\frac{\partial}{\partial r}\left(\frac{W}{r}\right)\right)^2+4\left(\frac{\partial \beta}{\partial r}+\frac{1}{r}\right)\frac{\partial \nu}{\partial r}.
    \end{split}
    \label{d}
\end{equation}
Here, $\pm$ corresponds to the prograde and retrograde orbits \cite{Grandclement:2014msa}, and the existence of the orbits is determined by the condition $D\geq 0$. $r_{ISCO}$, then, is the smallest radius such that real solutions exist.
We can also define the Keplerian angular momentum and energy with respect to a ZAMO (Zero Angular Momentum Observer) as
\begin{equation}
   \begin{split}
       &l=\frac{e^{\beta 
       }rv(r)}{e^{\nu}-e^{\beta}Wv(r)
       },\\&
       E=\frac{e^{\nu}-e^{\beta}Wv(r)}{\sqrt{1-v(r)^2}} .
   \end{split} 
\end{equation}

\begin{figure*}[]
%\vspace*{-1cm}
\centering
\hspace*{-0.7cm}\includegraphics[width=0.50\textwidth]{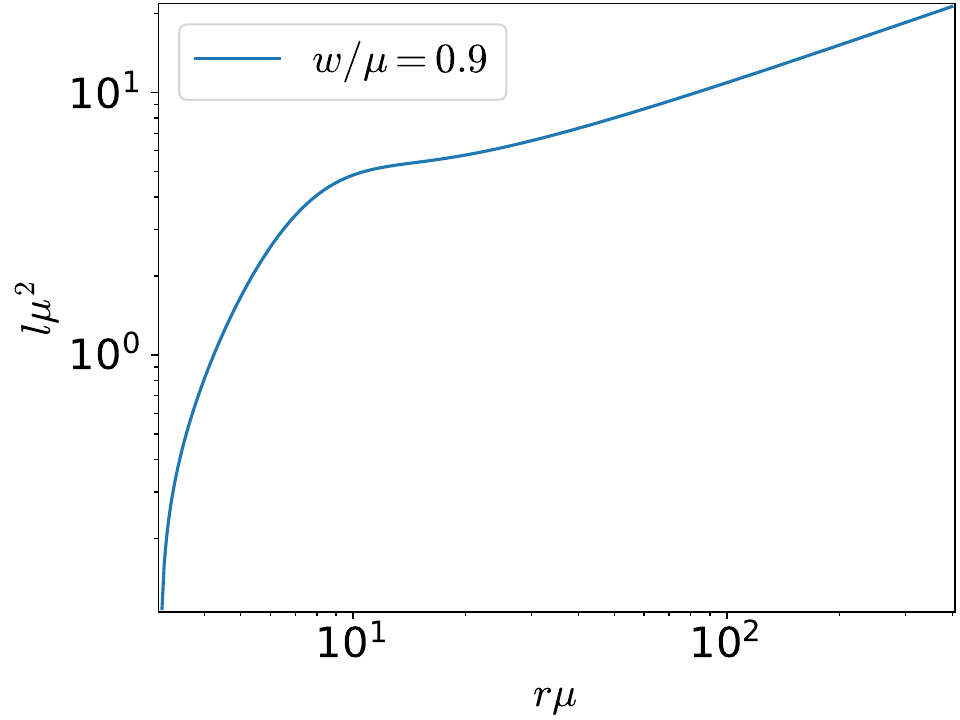}
\caption{For the BS with non-self-interacting potential, and $w/\mu=0.9$ we show the values for the angular momentum $l$ in terms of the radial coordinate. We can read the same range and behavior as in \cite{Meliani:2015zta} for the same model.}
\label{lrplot}
\end{figure*}

For a scalar BS with $w=0.9$, we plot $r-l$, which allows us to compare our results \cref{lrplot} with fig. 2 in  \cite{Meliani:2015zta}, with an excellent agreement.

Based on the desired metric functions for a specific model, we derived an equation describing the velocity of matter orbiting around the source. This formulation is typically used to study orbits in stellar systems. BS and PS have sizes comparable to, or within the same order of magnitude as, conventional Neutron Stars (NS). Our study treats the EKG and EP systems as galactic halos. By considering the boson mass as a free parameter and recognizing that it rescales the problem, we can employ realistic values to obtain solutions that range from the sizes of NS to entire galaxies. The specific parameters and scaling factors are detailed in \cref{rescalingsect}. Consequently, we can model scalar and vector bosonic candidates for dark matter halos.

Next, we have to obtain the masses for the EKG and EP systems in units of the boson mass.
After solving for the metric functions, the mass can be extracted in terms of the Komar integrals through the stress-energy tensor for each system \cite{stergioulas1994comparing},
\begin{equation}
\begin{split}
    & M_{(A,\Phi)}=\int (-2T_{(A,\Phi)t}^t+T_{(A,\Phi)\sigma}^{\sigma})\sqrt{-g}d^3x.\\
    \end{split}
    \label{masss}
\end{equation}
The sub-labels $A$ and $\Phi$ refer to the EP and EKG cases, respectively. We can easily obtain the mass by furnishing the equation with the correspondent Stress-Energy tensor components.
We have to take into account that we obtain the quantities $ M\mu$ rescaled by the boson mass, which must be converted to dimensional units when dealing with real data.

%%%%%%%%%%%%%%%%%%%%%%%%%%%%%%%%%%%%%%%%%%%%%%%%%%%%%
\subsection{Curves for distinct scalar models}
\label{curvesmodelsbs}
%%%%%%%%%%%%%%%%%%%%%%%%%%%%%%%%%%%%%%%%%%%%%%%%%%%%
We have studied the rotation curves for four massive models of scalar BS. 
We do not consider the case of a spinning, massless boson here. The main reason is that we use the boson mass for the rescaling in our numerical integration, and to obtain massless solutions would require a complex restructuring of the code. But we want to mention that a massless potential has been studied in the static BS framework as a candidate for galactic halos in \cite{Schunck:1998nq}. The first model shown is the usual non-self-interacting potential, in the non-rotating case also called the \textit{Mini BS}. This model is very commonly used in the literature \cite{Schunck:2003kk,Liebling:2012fv,Lai:2004fw},
\begin{equation}
V_{Mini}=\mu^2\phi^2.
 \end{equation}
We also examine the negative self-interacting model, a specific variant of the quartic model characterized by a negative coupling constant. Negative coupling constants turn out to lead to lower velocities, more in line with the observed galactic rotation curves. We will see, however, that this reduction is insufficient. We refer to the potential with an \textit{H} (=halo),

\begin{equation}
V_{H}=\mu^2\phi^2-\alpha\phi^4
\begin{cases}
      \alpha= 1,\\
      \alpha=12,\\
      \alpha=50.
\end{cases}
 \end{equation}

The axion BS is also a prototypical kind of bosonic matter. 
We decided to study its velocity curves because, depending on the coupling parameters \cite{Delgado:2020udb},
very distinct types of solutions can be reached.
%Due to the properties of this potential, reaching very distinct values for the solutions depending on the coupling parameters \cite{Delgado:2020udb}, we decide to study their velocity curves as a candidate. 
 \begin{equation}
\begin{split}
V_{ Ax}&=\frac{2f^2\mu^2}{B}\left(1-\sqrt{1-4B\sin^2(\frac{\phi}{2f})}\right)\\&
\begin{cases}
      f=0.005, &B=0.22.
\end{cases}
\end{split}
 \end{equation}
The last case we consider is the so-called solitonic potential. This potential was proposed and studied as a valuable candidate for dark matter in the static framework \cite{Lee:1995af,Schunck:2003kk},
\begin{equation}
V_{Sol}=\mu^2\phi^2\left(1-\left(\frac{\phi^2}{\phi_0^2}\right)\right)^2
\begin{cases}
      \phi_0=0.05, \\
\end{cases}
 \end{equation}  

This last case can be interpreted, to some extent, as a limit of the axionic potential. However, in this study, we consider both this leading-order limit and the case of the full axionic potential without approximation. Concretely, we have independently solved and analyzed the case without self-interaction, the negative quartic case (called halo), the full axionic case, and the solitonic case.

\begin{figure*}[]
%\vspace*{-1cm}
\centering
\hspace*{-0.7cm}\includegraphics[width=0.50\textwidth]{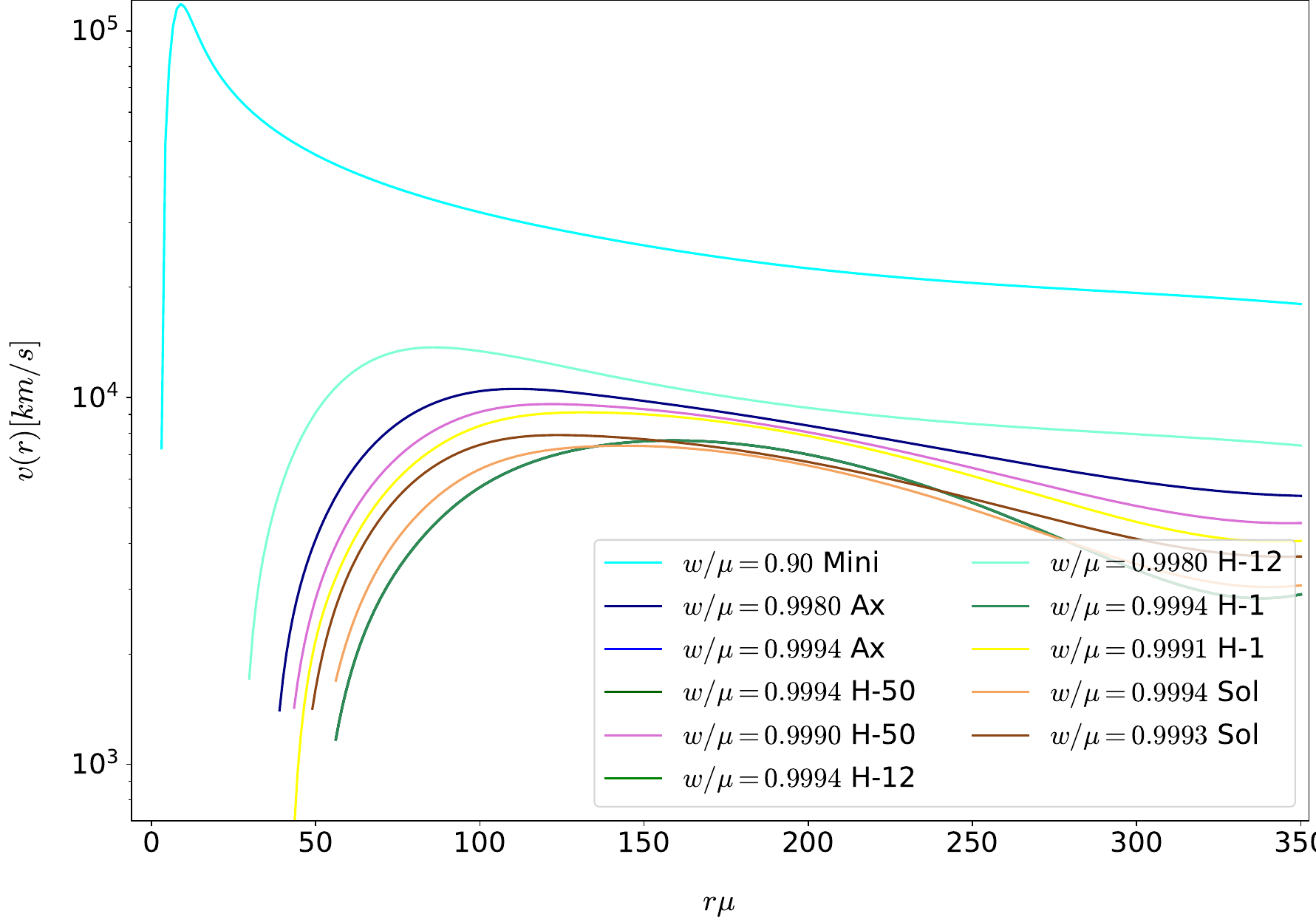}
\caption{Rotation curves for a set of selected models. The labels indicate the kind of self-interacting potential $V_{Mini},V_{H},V_{Ax}$ and $V_{sol}$. For the $V_H$, we used three different coupling constants: $1,12$ and $50$. For all the cases but the $Mini$, we show two different curves based on the internal field frequency $w/\mu$.}
\label{bscurvespots}
\end{figure*}

Specifically, we obtained the rotation curves for internal field frequencies in the range $w/\mu\in [0.9,1)$ for $V_{Mini}$ and $V_H$, whereas for the solitonic and axionic potentials we consider the larger interval $w/\mu\in \left[0.28,1\right)$. We show in \cref{bscurvespots}  a generic set of solutions, and express the velocities in the usual astrophysical system of units, $\mathrm{[km/s]}$. It is clearly seen that all cases exceed $10^3\mathrm{[km/s]}$. The known galaxies, specifically the ones selected for our study, even in the limiting case have maximum rotation velocities below $300\mathrm{[km/s]}$. This implies that models of rotating scalar BS lead to excessive velocities and cannot be used to fit the galactic dark matter halos. 
%In more detail, looking at \cref{bscurvespots}, we  see that the behavior is expected, and the shape of the velocity profiles for this system could agree with the observations. Still, the high velocities deny the possibility of taking them into account. 
It is important to remark that we have found the curves in various different regimes. The best region in the internal frequency parameter was shown to be $w/\mu\in[0.9980,0.9994]$, which is the non-relativistic region. Still, even in this extreme regime, near the limit of the existence of solutions $w/\mu=1$, the velocities remain too high for a correct and realistic fitting. 
It is interesting to mention that the static case, as shown in the cited bibliography, was considered due to its good properties, but a major difference splits the static and the spinning cases. Rotating scalar bosonic fields coupled to gravity spread in space-time with a toroidal rather than spherical matter distribution. This toroidal nature makes these objects show higher inertial and quadrupolar properties, which can be related to these higher velocities.

We conclude this subsection by confirming that the rotating scalar boson systems considered here are unable to describe galactic halos as dark matter.

%%%%%%%%%%%%%%%%%%%%%%%%%%%%%%%%%%%%%%%%%%%%%%%%%%%%%%%%
\subsection{Curves for the Proca models}
\label{curvespsmodel}
%%%%%%%%%%%%%%%%%%%%%%%%%%%%%%%%%%%%%%%%%%%%%%%%%%%%%%%%%%%%

In contrast to the scalar stars, the Proca framework maintains similar matter distributions for the static and spinning situations. Even in the Proca case the two situations are not continuously connected because of the quantization of the angular momentum, but both static and spinning Proca stars are deformed spheroids.  Moreover, as shown in \cite{Adam:2024zqr}, these systems' inertial and quadrupolar momenta are smaller than for their scalar counterparts. 
Further, the Proca model has a particular issue related to the role of self-interactions. While Proca models with self-interactions are theoretically well-proposed, some instabilities and hyperbolicity problems arise, and their study is beyond the purposes of the present paper \cite{Coates:2022nif,PhysRevLett.129.151102,Mou:2022hqb,PhysRevLett.129.151103,Barausse:2022rvg,Brito:2024biy}.

We have studied Proca models in the frequency range $w/\mu\in[0.9955,1)$, using each frequency as a model parameter with a given name referring to the two last numbers of the frequency, going from $m55$ in the lower case to $m10$ for the limiting situation.  All the model velocities are obtained following the expression \cref{velocityhalo} using the corresponding data for each model.

\begin{figure*}[]
%\vspace*{-1cm}
\centering
\hspace*{-0.7cm}\includegraphics[width=0.50\textwidth]{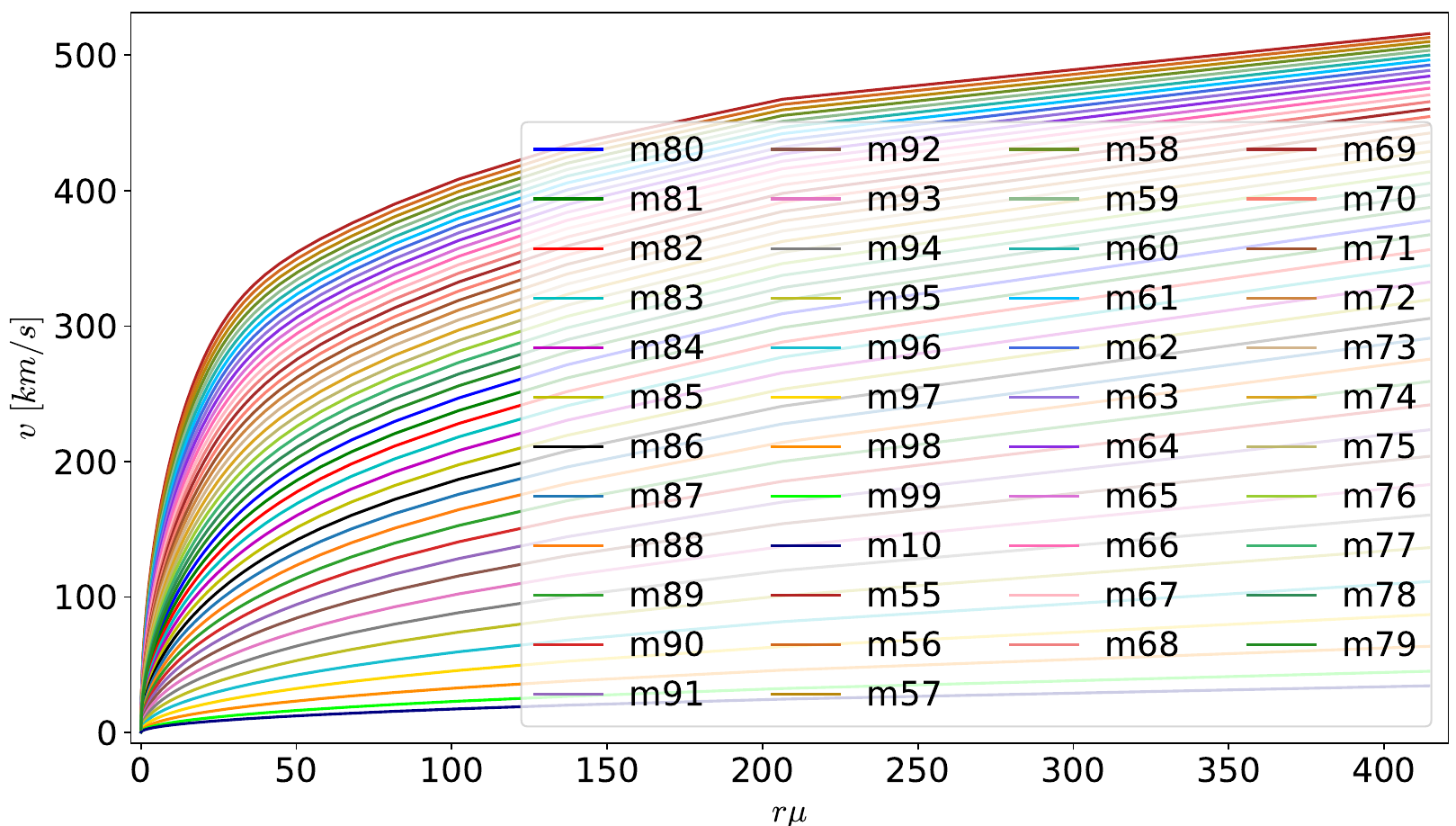}
\caption{Proca models for the galactic haloes. The radii are shown in units of the boson mass, while the velocities are shown in the usual units for the case $\mathrm{km/s}$. }
\label{procasmodelscurvessolos}
\end{figure*}

As depicted in \cref{procasmodelscurvessolos}, we can see that the soft behavior of the curves together with the maximal velocities ranging from less than $50 \mathrm{km/s}$ to  $\sim 400 \mathrm{km/s}$ allows us to choose the model that best suits each galaxy fitting case. Again, it is remarkable that the models with the best properties are those with field frequencies near the non-relativistic limit.

The main conclusion of this subsection is that through the correct family of values for the mass of the boson, which allows us to rescale the radial coordinate, we can use the EP system as a candidate for the dark matter halos in the galactic fitting context. 
It is also nice to
have the internal field frequency as the model parameter, because thus we do not have to change the kind of interaction to describe different galaxies. Another advantage of this model is that we have a velocity curve previous to the real data fitting, not vice versa. This makes our model very versatile and useful once we have the numerical machinery for solving the EP system.

%%%%%%%%%%%%%%%%%%%%%%%%%%%%%%%%%%%%%%%%
\section{Matter description and observable curves}
\label{models}
%%%%%%%%%%%%%%%%%%%%%%%%%%%%%%%%%%%%%%%%

As we show in this section, a prototypical galaxy comprises several key components: the stellar and gas disks, the bulge, and the halo. The bulge, located at the galaxy's center, is a dense, spheroidal region populated with older stars. Surrounding the bulge is the stellar disk, a flattened, rotating structure containing most of the galaxy's stars, including younger, more luminous ones. Intermixed with the stellar disk is the gas disk, composed mainly of hydrogen gas, which serves as the raw material for star formation. Enveloping these components is the halo, an extended, roughly spherical region containing older stars, globular clusters, and dark matter, which dominates the galaxy's mass and influences its gravitational dynamics \cite{binney2008galactic}.

\begin{table}
\begin{tabular}{|l|c|c|c|c|c|c|}
\hline
\textbf{galaxy} & $L$ & $R_d$ & $(M/L)d$ & $M_{HI}$ & $(M/L)b$ &$ HI/He$\\
\hline
DDO154 & 0.05 & 0.50 & 1.4 & 5.0 & 0 & 1.5   \\
DDO170 & 0.16 & 1.28 & 3.7 & 6.0 & 0 & 1.3   \\
NGC1560 & 0.35 & 1.30 & 3.0 & 3.5 & 0 & 1.6   \\
UGC2259 & 1.02 & 1.33 & 4.5 & 4.0 & 0 & 1.4   \\
NGC2403 & 7.90 & 2.05 & 2.00 & 43.0 & 0 & 1.2   \\
NGC2903 & 15.3 & 2.02 & 3.50 & 61.0 & 0 & 1.7   \\
NGC2998 & 12.0 & 5.40 & 0.25 & 20.0 & 1.5 & 1.5  \\
NGC3109 & 0.81 & 1.55 & 0.4 & 7.5 & 0 & 1.4  \\
NGC3198 & 9.00 & 2.63 & 3.83 & 37.0 & 0 & 1.6   \\
NGC7331 & 54.0 & 4.48 & 2.25 & 80.0 & 1.7 & 1.3   \\
\hline
\end{tabular}
\caption{We present the parameters for the galaxies, including luminosities $L$ in units of $\mathrm{10^9M_{\odot}}$, effective radii $R_d$ in $\mathrm{kpc}$, mass disk/luminosity ratio $(M/L)d$,  HI mass $M_{HI}$ in $\mathrm{10^6M_{\odot}}$, mass bulge/luminosity ratio $(M/L)b$ and the HI/He ratio. The numerical values for these quantities are derived from studies of mass distributions and the photometric and dynamical properties of spiral and dwarf galaxies \cite{Ashman1988, Blanton2013, Read2003, Trentham2003, Sofue1997, Elmegreen2016, Silk2002, 1996A&AS..118..557D,Kormendy2013,Schunck:1998nq}. We also use the value for the effective bulge radius for the galaxies \textit{NGC2998}, $\sim 0.45$, and for \textit{NGC7331}, $\sim1$. Further sources for the data are \cite{things2008,alfalfa2012,sparc2016,tng2024,aip2021,schombert2017}.}
\label{table1}
\end{table}

Consequently, the rotation curves for galaxies can be fitted by parts, considering separate contributions. Several distinct models exist for each velocity component, all based on observable quantities related to the measured luminosity of the galaxy under study. 
This work aims to show the possibility of vectorial bosons as dark matter halo particles. We use well-known models for the luminous part, the sum of the disk and bulge. 
The following section introduces the models for the different velocity components used in our research.

%%%%%%%%%%%%%%%%%%%%%%%%%%%%%%%%%%%%%%%%%%%%%%%%%%%%%%%%%%%%%%
\subsection{Disk}
%%%%%%%%%%%%%%%%%%%%%%%%%%%%%%%%%%%%%%%%%%%%%%%%%%%%%%%%%%%%%%
We split the disk contribution into two components: the disk velocity due to stellar matter and the disk velocity due to gas, as given by the HI (neutral hydrogen) measurements. 

For the stellar disk distribution, we use the thin disk approximation proposed by Freeman \cite{1970ApJ...160..811F} and following notation used in \cite{galacticDyn} ,
\begin{equation}
    v_d^2(r)=4\pi G\Sigma_0R_dy^2[I_0(y)K_0(y)-I_1(y)K_1(y)].
\end{equation}
Here, $\Sigma_0=M_d/(2\pi R_d^2)$, where $M_d$ is obtained through the multiplication of the total luminosity by the mass disk/luminosity ratio, with the corresponding units in solar masses.  $R_d$ is the scale radius of the disk, also a constant that is empirically adjusted for each galaxy.   $I_n(y)$ and $K_n(y)$ are the modified Bessel functions, with $n=0,1$, and $y$ is the following ratio of the radial coordinate over the effective radius,
\begin{equation}
    y=\frac{r}{2R_d}
\end{equation}

This approach for parametrizing the stellar disk matter is commonly used in the literature \cite{1970ApJ...160..811F,Courteau:1997ap,Schunck:1998nq,Sofue:2000jx}.

We obtain the gas contribution based on empirical observations and theoretical models that describe how the density of HI gas decreases exponentially with the radial distance $r$ from the galactic center \cite{binney2008galactic,bosma1981distribution,begeman1989hi,deblok2008high}.

The adjustment for the surface density of HI gas follows a decreasing exponential law, which is commonly used in galactic disk fittings. The specific formula used is:
\begin{equation}
 \Sigma_{\text{HI}}(r) = \Sigma_{0,\text{HI}} \cdot \exp\left(-\frac{r}{R_d}\right) 
\end{equation}

Here,
 $\Sigma_{\text{HI}}(r) $ is the surface density of HI gas at a radial distance $r$.
$\Sigma_{0,\text{HI}} $ is the central surface density of HI/He gas, a constant that depends on the specific galaxy and is obtained for each case by the product of HI mass given in \cref{table1} with the scale in solar masses; $R_d$ is the same scale radius as above.

The exponential density profile is a well-established astrophysics model for describing matter distribution in galactic disks. This model is derived from the solution to the equation of hydrostatic equilibrium in a thin, self-gravitating disk, and it also considers the dynamics of galaxy formation and evolution \cite{binney2008galactic}.

The velocity of HI gas is calculated by integrating the surface density of the gas over the radius. The circular velocity at a distance $r$ is given by:
\begin{equation}
    v_{\text{g}}(r) = \sqrt{ \frac{G M_{\text{gas}}(r)}{r}} ,
\end{equation}
where $  M_{\text{gas}}(r) $ is the mass of the gas within the radius $r$. This mass is obtained by integrating the surface density of the gas,
\begin{equation}
     M_{\text{gas}}(r) = \int_0^r 2 \pi r'  \Sigma_{\text{HI}}(r') dr' .
\end{equation}

%%%%%%%%%%%%%%%%%%%%%%%%%%%%%%%%%%%%%%%%%%%%%%%%%%%%%%%%%%%%%%
\subsection{Bulge}
%%%%%%%%%%%%%%%%%%%%%%%%%%%%%%%%%%%%%%%%%%%%%%%%%%%%%%%%%%%%%%
We use the Sérsic parametrization for the bulge, which is a generalization of the de Vaucouleurs' function
\begin{equation}
    I(r)=I_e e^{-b_n\left[\left(\frac{r}{R_{eb}}\right)^{\frac{1}{n}}-1\right]}
\end{equation}
with $b_n=7,6692$, $n=4$ and $R_{eb}$ the effective bulge radius, and
\begin{equation}
  I_e=L_{bulge}\frac{b_n^{2n}}{2\pi n e^{b_n}\Gamma(2 n)R_{eb}^2}  
\end{equation}
 with $L_{bulge}$ the fraction of the luminosity due to the bulge \cite{Sofue:2008wt,Caon:1993wb} and $\Gamma(x)$ the Euler gamma function. Velocities due to a mass distribution following this kind of profile, do not have a simple analytical solution, so we numerically integrate it,
\begin{equation}
    M_b(r)=\int_0^r 2\pi r' I(r')dr'.
\end{equation}

The circular velocity in a bulge with our de Vaucouleurs profile is
\begin{equation}
    v_b(r)=\sqrt{\frac{GM_b(r)}{r}} .
\end{equation}

%%%%%%%%%%%%%%%%%%%%%%%%%%%%%%%%%%%%%%%%
\subsection{Proca stars as DM}
%%%%%%%%%%%%%%%%%%%%%%%%%%%%%%%%%%%%%%%%
To parametrise the halo component, we have derived the velocity of the orbits produced by axisymmetric EKG and EP systems \cref{velocityhalo} in the limit of $w/\mu\rightarrow 1$, which means that we are using only models in the non-relativistic limits. Otherwise, the velocities for the bosonic parts take too high values. This is aligned with the fuzzy DM models often used in this context \cite{Painter:2024rnc,Bullock:2017xww,DelPopolo:2016emo}.

As explained in \cref{curvespsmodel}, we can examine the shape of the velocity curve components for distinct models of EP systems. All the models shown in \cref{procasmodelscurvessolos} present a soft behavior with the correct range of velocities. Fixing the boson mass to a specific value $\mu\sim 10^{-26}\mathrm{eV}$ allows the halo to have the right length scale. The galaxies under study range from $8$ to $40$ $\mathrm{kpc}$, corresponding to $\mu=34\cdot 10^{-26}\mathrm{eV}$ and $\mu=6\cdot 10^{-26}\mathrm{eV}$, respectively.  Still, we have to perform a data interpolation. After fixing $\mu$, we obtain radii in the correct order of magnitude. The second parameter of the model is the internal field frequency. Our halo models depend on $\mu$ and $w/\mu$, resulting in rotation components from the EP systems as $v_P(r)=\bar{v}_P^{\mu,w}(r)$, allowing us to select the best option for plausible candidates of DM halo curves in each case.

%%%%%%%%%%%%%%%%%%%%%%%%%%%%%%%%%%%%%%%%
\subsection{Total rotation curves}
%%%%%%%%%%%%%%%%%%%%%%%%%%%%%%%%%%%%%%%%

\begin{figure*}[]
%\vspace*{-1cm}
\centering
\hspace*{-0.7cm}\includegraphics[width=0.50\textwidth]{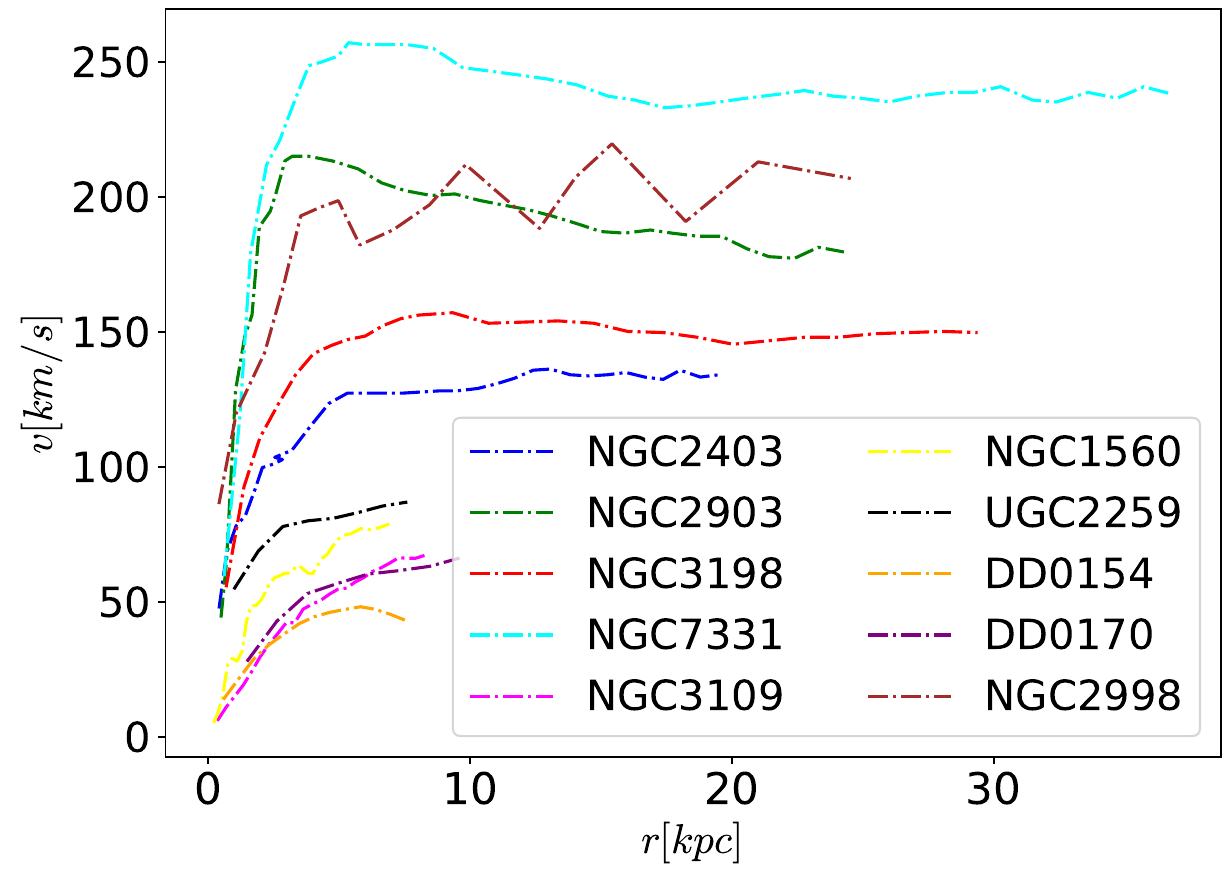}
\caption{Observational rotation curves for the galaxies we study in this paper \cite{Schunck:1998nq}.}
\label{datareal}
\end{figure*}

Our fitting models for the rotation curves have three or four components, depending on whether the bulge is present in a given galaxy.  Two of them fit the luminous matter as a disk, one a possible bulge, and the remaining one describes the contribution of a halo produced by a complex vectorial field coupled to gravity in the non-relativistic limit.
The total velocity is given by
\begin{equation}
    v_T^2(r)=v_d^2(r)+v_g^2(r)+v_b^2(r)+v_P^2(r)
\end{equation}

The process of fitting to the real data is done by inspection. Once the luminous contributions and the halo length scale are fixed, we choose the frequency model that is  best for each galaxy and try for a given $w$ to fit $v_T^2$.

We will compare our models with the galaxies treated in \cite{Schunck:1998nq}. As we can read from  \cref{datareal}, the ten presented galaxies are very different in sizes and maximal velocities, constituting a representative set of cases.

%%%%%%%%%%%%%%%%%%%%%%%%%%%%%%%%%%%%%%%%%%%%%%%%%%%%%%%%%%%%%%
\section{Results}\label{results}
%%%%%%%%%%%%%%%%%%%%%%%%%%%%%%%%%%%%%%%%%%%%%%%%%%%%%%%%%%%%%%

As shown in the previous section, we have observational data for ten galaxies. Now, we put together the real data with the luminous and DM parts for each galaxy. We indicate $\mu$ and $w/\mu$ for each model, together with the fittings. The observed velocity data are represented by black diamonds in all figures.
%%%%%%%%%%%%%%%%%%%%%%%%%%%%%%%%%%%%%%%%%%%%%%%%%%%%%%%%%%%%%%
\subsection{Fittings with the EP models}\label{results1}
%%%%%%%%%%%%%%%%%%%%%%%%%%%%%%%%%%%%%%%%%%%%%%%%%%%%%%%%%%%%%%
As detailed in the previous section, we have compiled observational data for ten galaxies. We have now integrated this data with models encompassing each galaxy's luminous components and dark matter. We present key parameters for each model, including the boson mass ($\mu$) and the internal field frequency ($w/\mu$), alongside the model fittings. These parameters and their fittings help illustrate the accuracy and efficacy of our models in representing real galactic structures.

\begin{figure*}[]
%\vspace*{-1cm}
\centering
\hspace*{-0.7cm}\includegraphics[width=0.50\textwidth]{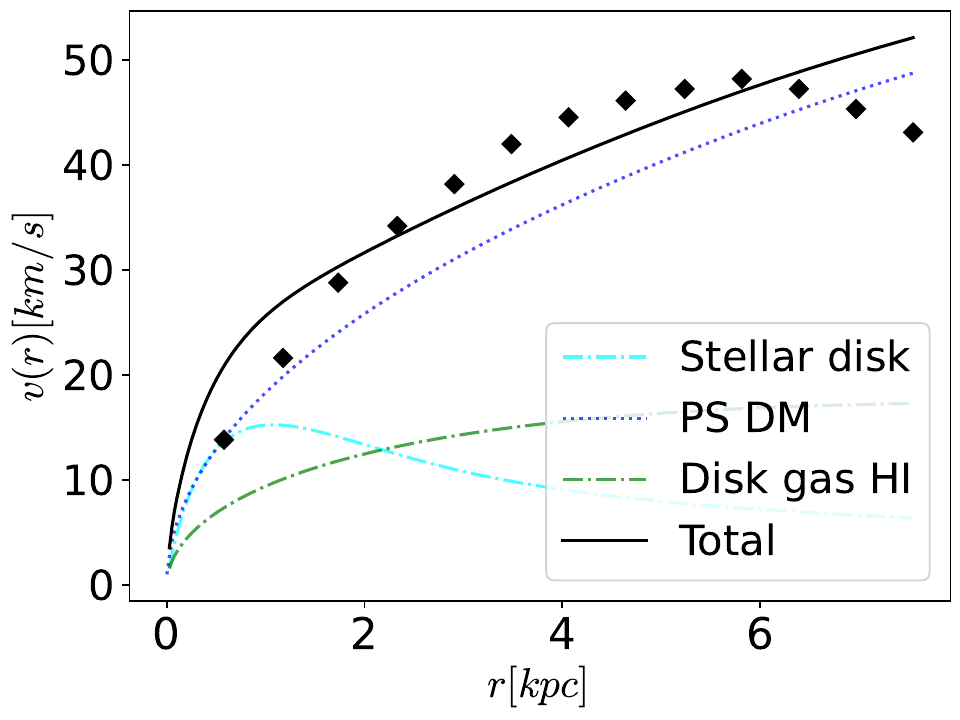}
\caption{Rotation Curve fitting for\textit{DDO 154}.}
\label{DDO154}
\end{figure*}

Our analysis begins with \textit{DDO 154}, a gas-rich, irregular dwarf galaxy known for its significant dark matter halo, as highlighted in several studies \cite{1988ApJ...332L..33C}. In terms of our modeling, the best fit was achieved with a boson mass, $\mu$, set at $25\cdot 10^{-26}\mathrm{eV}$ and a field frequency, $w/\mu$, of $0.9998$. This model successfully approximates the galaxy's radius at approximately $8 \mathrm{kpc}$ and maintains velocity values under $50\mathrm{km/s}$. While the fit closely aligns with the data \cref{DDO154}, it suggests the possibility of additional dark matter contributions not accounted for in our current model. Nevertheless, the overall shape and magnitude of the velocity curve produced by our model are deemed satisfactory.

\begin{figure*}[]
%\vspace*{-1cm}
\centering
\hspace*{-0.7cm}\includegraphics[width=0.50\textwidth]{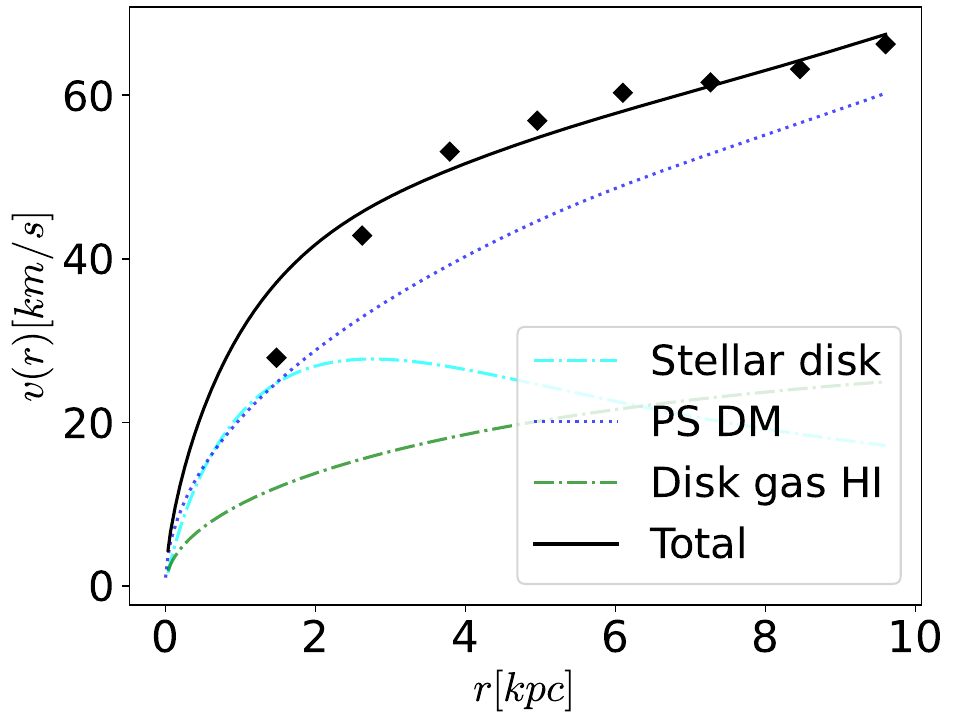}
\caption{Rotation Curve fitting for \textit{DDO 170}.}
\label{DDO170}
\end{figure*}

The next case is the dwarf galaxy \textit{DDO 170}, as documented by \cite{1990AJ.....99..547L}. Similar to \textit{DDO 154}, this galaxy is influenced by its DM content but exhibits some unique characteristics in its dynamic behavior. In modeling the effects of the halo, we set the boson mass, $\mu$, at $23\cdot 10^{-26}\mathrm{eV}$ and the field frequency, $w/\mu$, at $0.9998$. This results in a velocity curve that is faster yet roughly equivalent in size to that of \textit{DDO 154}, indicating a strong similarity in the DM distribution between the two galaxies. Our model's fitting closely matches the observational data from \cref{DDO170}, suggesting a high accuracy in representing the galaxy's dynamics under the influence of its dark matter halo.

\begin{figure*}[]
%\vspace*{-1cm}
\centering
\hspace*{-0.7cm}\includegraphics[width=0.50\textwidth]{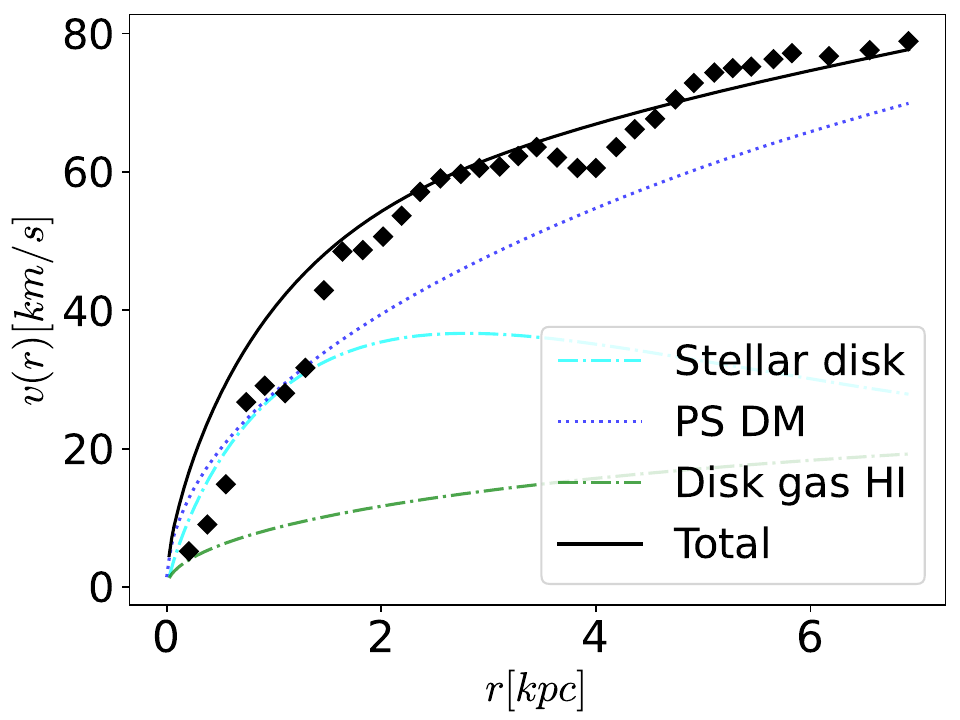}
\caption{Rotation Curve fitting for \textit{NGC 1560}.}
\label{NGC1560}
\end{figure*}

The third galaxy is the spiral galaxy \textit{NGC 1560}, located in the \textit{Camelopardalis} constellation. This galaxy mirrors characteristics observed in the previous cases, notably a diameter less than $10\mathrm{kpc}$ and the absence of a significant bulge. For our dark matter modeling, the boson mass was set at $\mu=24\cdot 10^{-26}\mathrm{eV}$ and the field frequency at $w/\mu=0.9997$. These parameters yielded highly accurate results. As demonstrated in \cref{NGC1560}, the aggregate of the contributions produces a velocity curve that closely aligns with the observational data, affirming our model as a strong contender for elucidating the role of DM in this galaxy.

%The \textit{Perseus} galaxy \textit{UGC2259} is a dwarf type studied in  \cite{1991A&A...242..334G}. With observational data for radii below $8\mathrm{kpc}$ and maximal velocities surpassing $80 \mathrm{km/s}$, the DM parameters that allow the best fitting are $\mu=34\cdot 10^{-26}\mathrm{eV}$ and $w/\mu=0.9998$. Again, our model fits quite well with reality, having, in general, the correct behavior and quantities \cref{UGC2259}. Together with the previous dwarf galaxies, this non-bulge galaxy fits considerably well our assumption of vectorial bosonic DM.

\begin{figure*}[]
%\vspace*{-1cm}
\centering
\hspace*{-0.7cm}\includegraphics[width=0.50\textwidth]{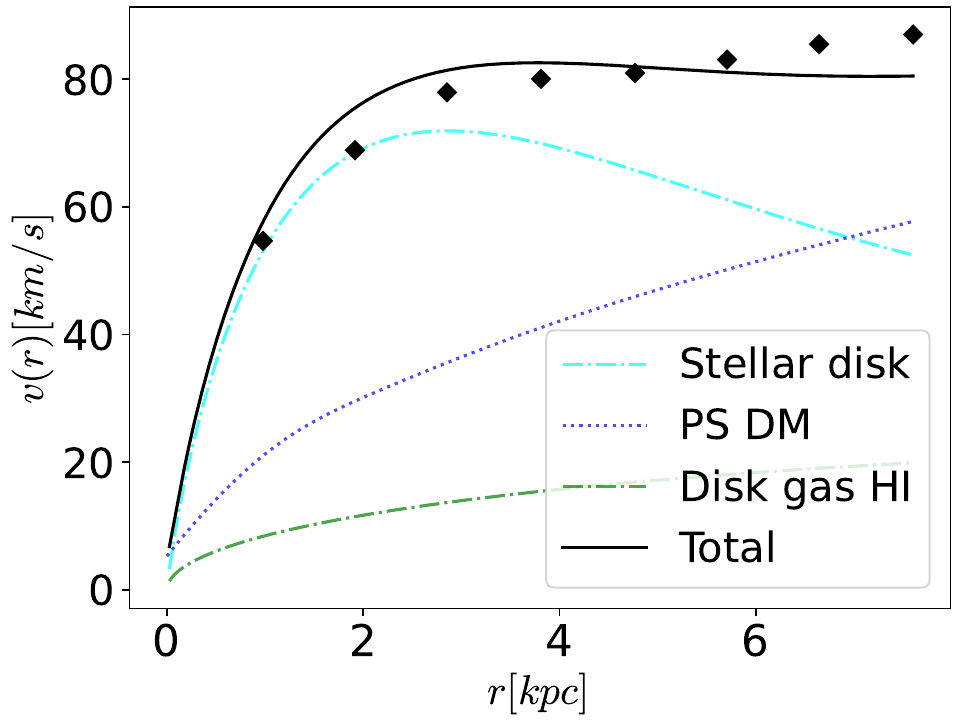}
\caption{Rotation Curve fitting for \textit{UGC 2259}.}
\label{UGC2259}
\end{figure*}

The dwarf galaxy \textit{UGC 2259} in the \textit{Perseus} constellation has been extensively studied, notably in \cite{1991A&A...242..334G}. Observational data indicate a diameter smaller than $8\mathrm{kpc}$ and maximum velocities exceeding $80\mathrm{km/s}$. For this galaxy, the optimal parameters were identified as $\mu$ of $34\cdot 10^{-26}\mathrm{eV}$ and a field frequency ($w/\mu$) of $0.9998$. These parameters resulted in a model that closely matches the observed data, as demonstrated in \cref{UGC2259}. This model exhibits the correct behaviors and magnitudes and confirms the effectiveness of our vectorial bosonic DM hypothesis.

\begin{figure*}[]
%\vspace*{-1cm}
\centering
\hspace*{-0.7cm}\includegraphics[width=0.50\textwidth]{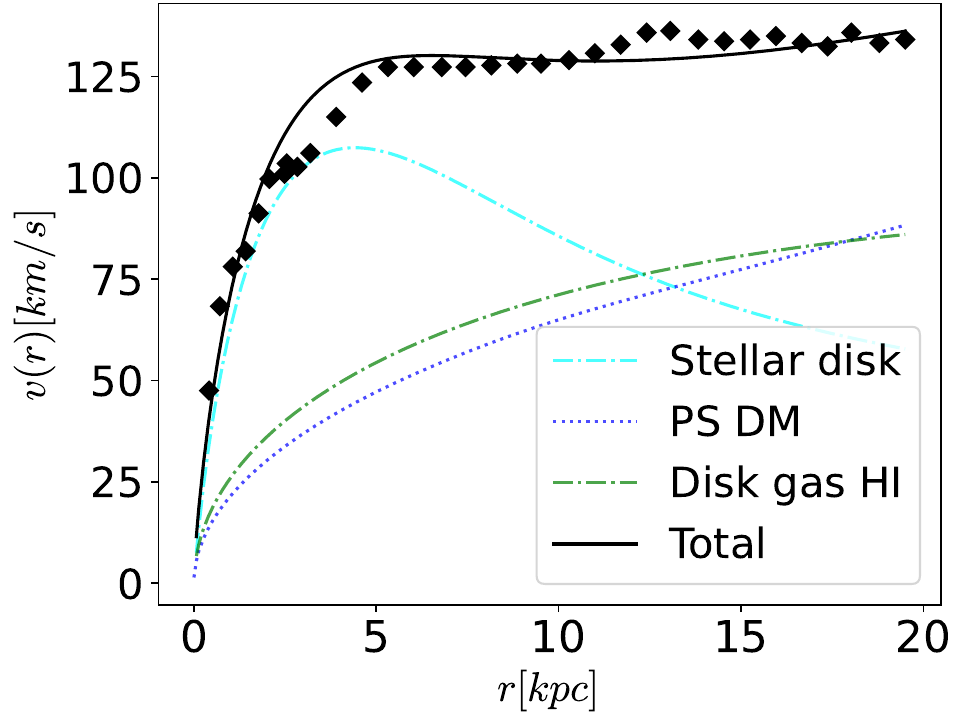}
\caption{Rotation Curve fitting for \textit{NGC 2403}.}
\label{NGC2403}
\end{figure*}

In the \textit{Camelopardalis} constellation, we also examine the spiral galaxy \textit{NGC 2403}. This galaxy stands out among the \textit{Global Cluster} galaxies as one of the nearest spiral galaxies to the \textit{Milky Way}, at a distance of approximately eight million light-years. The observational data reveal a rotation curve extending out to about $20\mathrm{kpc}$ with peak velocities exceeding $125 \mathrm{km/s}$. The optimal model for \textit{NGC 2403} was achieved with a boson mass of $14\cdot 10^{-26}\mathrm{eV}$ and a field frequency of $0.9997$. According to \cref{NGC2403}, our model impressively aligns with these observations, demonstrating its robustness in fitting the larger scale and higher velocity dynamics observed in this galaxy.

\begin{figure*}[]
%\vspace*{-1cm}
\centering
\hspace*{-0.7cm}\includegraphics[width=0.50\textwidth]{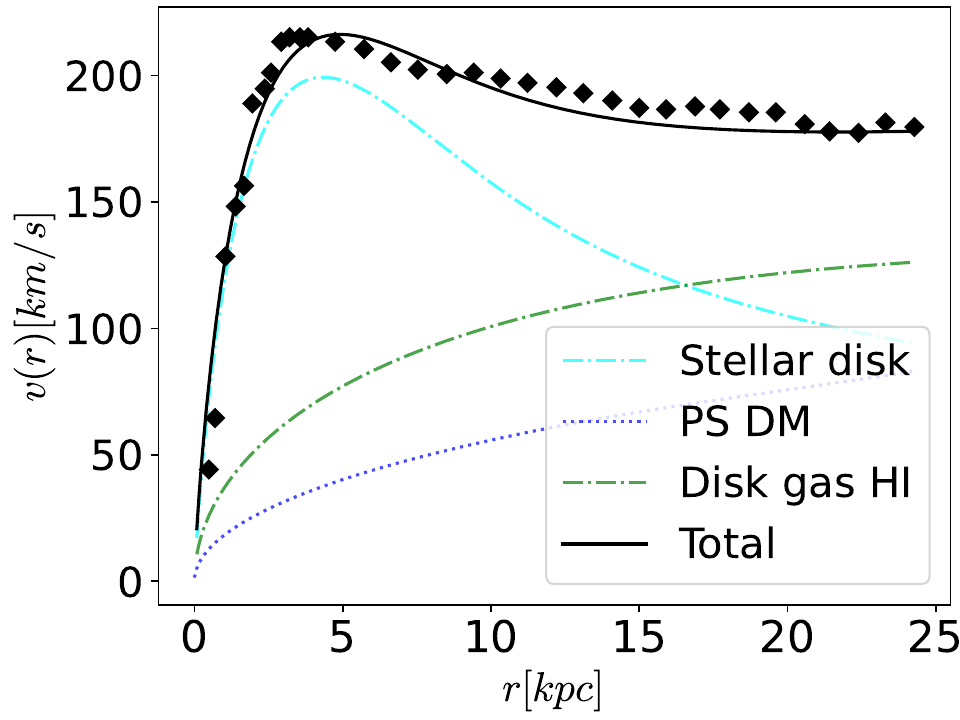}
\caption{Rotation Curve fitting for \textit{NGC 2903}.}
\label{NGC2903}
\end{figure*}

Located in the \textit{Leo} constellation, approximately twenty million light years from the \textit{Milky Way}, the bright spiral galaxy \textit{NGC 2903} shares several properties with our own galaxy. As detailed in \cref{NGC2903}, the observational data for \textit{NGC 2903} includes a rotation curve extending roughly $25\mathrm{kpc}$, with peak velocities exceeding $200\mathrm{km/s}$. For this galaxy, our EP model for the halo has proven to be remarkably precise. The model parameters are $\mu=10\cdot 10^{-26}\mathrm{eV}$ and a field frequency $w/\mu=0.9997$, which accurately capture this galaxy's dynamic characteristics and scale.

\begin{figure*}[]
%\vspace*{-1cm}
\centering
\hspace*{-0.7cm}\includegraphics[width=0.50\textwidth]{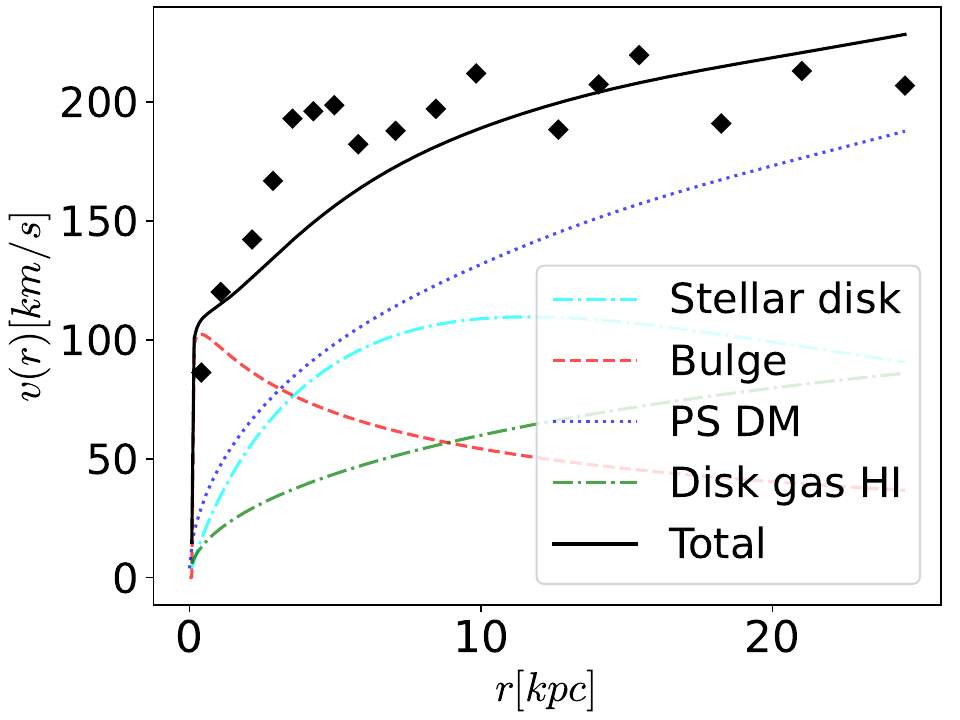}
\caption{Rotation Curve fitting for \textit{NGC 2998}.}
\label{NGC2998}
\end{figure*}

\textit{NGC 2998}, a barred spiral galaxy, is located approximately one hundred and ninety-five million light years from the \textit{Milky Way} in the \textit{Ursa Major} constellation. This galaxy is distinctive in our study from the previous cases for its bulge contribution to the velocity profile, as detailed in \cref{NGC2998}. The fitting plot reveals that our current model approximates the observed data closely but is not perfectly accurate. Additional dark matter components could enhance the velocity predictions in the $0$ to $15$ $\mathrm{kpc}$ region. Despite these challenges, the best model parameters found, $\mu=9\cdot 10^{-26}\mathrm{eV}$ and $w/\mu=0.9992$, align closely in both shape and magnitude with the empirical data.

\begin{figure*}[]
%\vspace*{-1cm}
\centering
\hspace*{-0.7cm}\includegraphics[width=0.50\textwidth]{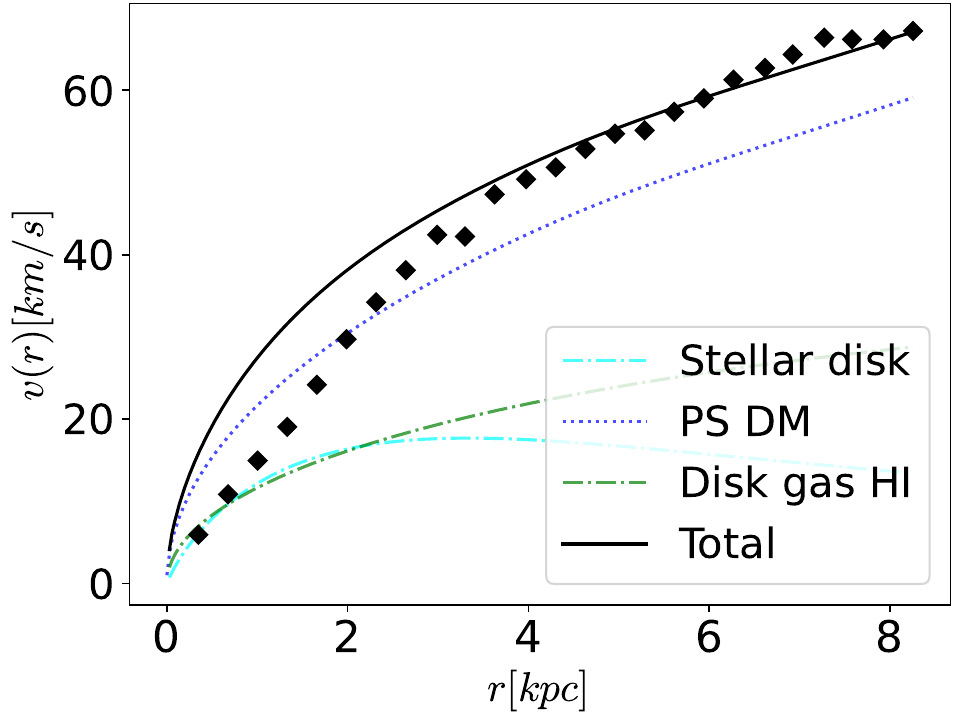}
\caption{Rotation Curve fitting for \textit{NGC 3109}.}
\label{NGC3109}
\end{figure*}

\textit{NGC 3109} is a compact barred galaxy of the Magellanic type, classified as either a spiral or irregular in structure. It is approximately $4.35$ million light-years away, in the direction of the \textit{Hydra} constellation. This galaxy is believed to be undergoing tidal interactions with the \textit{Antlia Dwarf}, a nearby dwarf elliptical galaxy. The parameters for our model, detailed in \cref{NGC3109}, were set at $\mu=28\cdot 10^{-26}\mathrm{eV}$ and $w/\mu=0.9998$. Notably, our model's fit tends to exceed the observational data near the galaxy's central region, a pattern observed similarly in previous cases involving smaller galaxies with moderate velocities.

\begin{figure*}[]
%\vspace*{-1cm}
\centering
\hspace*{-0.7cm}\includegraphics[width=0.50\textwidth]{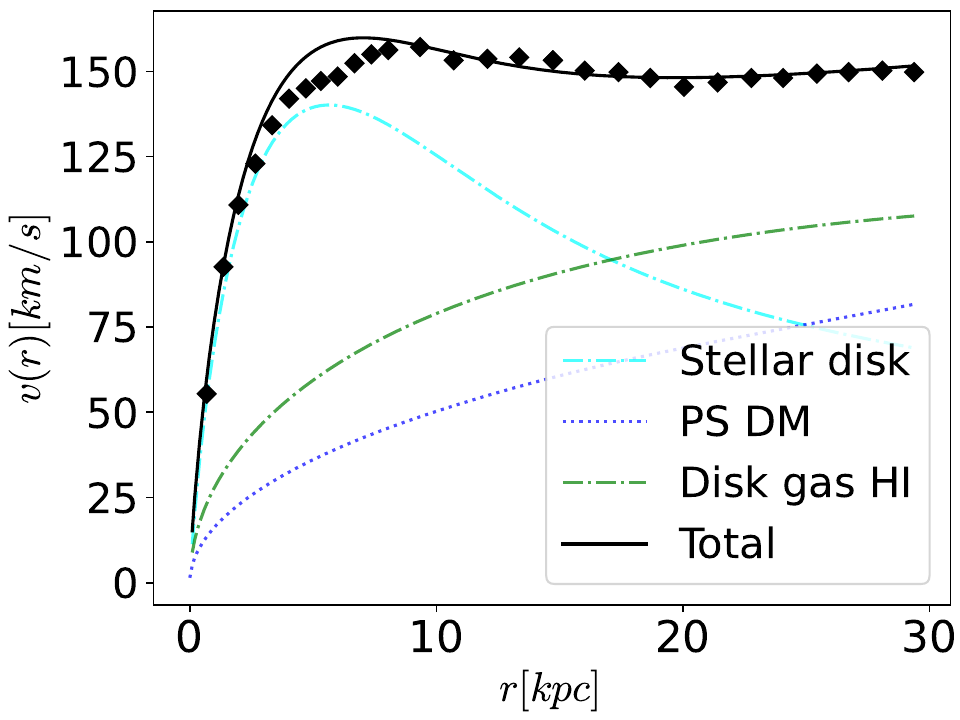}
\caption{Rotation Curve fitting for \textit{NGC 3198}.}
\label{NGC3198}
\end{figure*}

\textit{NGC 3198}, also cataloged as \textit{Herschel 146}, is a barred spiral galaxy located in the constellation \textit{Ursa Major}. Positioned about forty-seven million light-years away, it lies within the \textit{Leo Spur} of the \textit{Virgo Supercluster}. This galaxy played a significant role in the Hubble Space Telescope's Key Project on the Extragalactic Distance Scale, which aimed to refine the Hubble constant to within an accuracy of $10\%$.
Our model for \textit{NGC 3198}, presented in \cref{NGC3198}, uses parameters $\mu=8\cdot 10^{-26}\mathrm{eV}$ and $w/\mu=0.9997$. The fitting reveals that the galaxy maintains a rotation curve extending approximately $30\mathrm{kpc}$ with velocities surpassing $150 \mathrm{km/s}$. Remarkably, our model shows extreme accordance with the observational data, accurately capturing the dynamic properties of this galaxy.

\begin{figure*}[]
%\vspace*{-1cm}
\centering
\hspace*{-0.7cm}\includegraphics[width=0.50\textwidth]{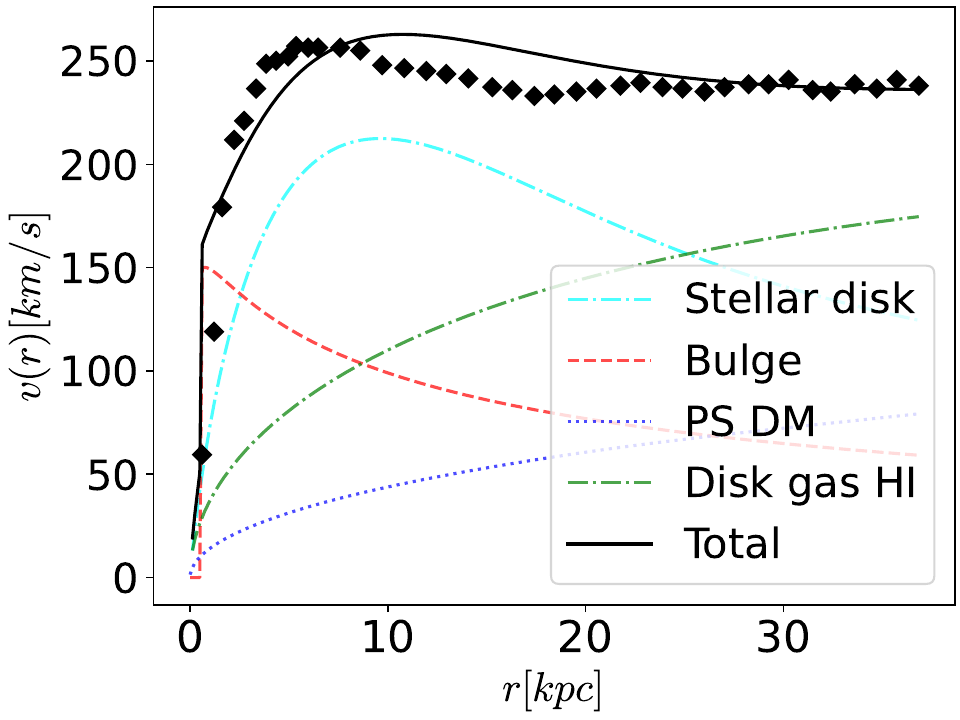}
\caption{Rotation Curve fitting for NGC7331 . EP m91}
\label{NGC7331}
\end{figure*}

The last galaxy in our study, \textit{NGC 7331}, also known as \textit{Caldwell 30}, is an unbarred spiral galaxy located approximately forty million light-years away in the constellation \textit{Pegasus}. It serves as the principal galaxy in the \textit{NGC 7331} Group, and is notably more prominent than its distant companions—\textit{NGC 7335, 7336, 7337}, and \textit{7340}—which are located about three hundred to three hundred fifty million light-years away \cite{Smith2021}. \textit{NGC 7331} also features a bulge that influences its gravitational dynamics.
As shown in \cref{NGC7331}, our model fitting closely aligns with the observational data, particularly in the region extending from $20$ to $40$ $\mathrm{kpc}$. This galaxy displays the largest radial extent and highest velocities observed in our study, with a radius exceeding $40$ $\mathrm{kpc}$ and velocities surpassing $250$ $\mathrm{km/s}$. The parameters used in our model are $\mu=6\cdot 10^{-26}\mathrm{eV}$ and $w/\mu=0.9997$, which have proven effective in reproducing the galaxy's dynamic properties.

After examining the cases on an individual basis, we can draw some overarching conclusions and offer specific insights regarding certain galaxies. Let's begin with the general findings. As observed, all EP models for DM galactic halos ensure that the total velocities—calculated as the square root of the sum of the luminous and dark matter components—align well with the experimental RCs. This consistency validates our methodological approach. Our analysis also reveals a central value for the boson's mass, consistently in the ultralight range, averaging around $\mu=18\cdot 10^{-26}\mathrm{eV}$, with variations ranging from $6$ to $34$ $\mathrm{eV}$. Although the range of these values is considerable, it is important to understand that the selection of a specific mass value is influenced by the diversity in the experimental measurements' scales, as discussed in Section \ref{rescalingsect}. Here, $\mu$ is adjusted to match the scale pertinent to each galaxy studied. Furthermore, all these values fall within the same order of magnitude, indicating a robust model consistency.

Moreover, the most accurate fits yielded frequencies of $ w/\mu=0.9997$ and $ w/\mu=0.9998$, suggesting a dominant frequency in the field.

Our analysis shows that larger galaxies with higher velocities generally demonstrate better compatibility with our EP models. This is particularly evident in the cases of \textit{NGC 2998},  the cases of \textit{NGC 7331}, \textit{NGC 2403}, \textit{NGC 2903} and \textit{NGC 3198}, which align exceptionally well with our EP dark matter halo model. The only exception is \textit{NGC 2998}, where our model appears nearly correct but lacks an additional dark matter component.

%%%%%%%%%%%%%%%%%%%%%%%%%%%%%%%%%%%%%%%%%%%%%%%%%%%%%%%%%%%%%%
\subsection{Fittings with the EP plus an Extra Dark Matter model}\label{results2}
%%%%%%%%%%%%%%%%%%%%%%%%%%%%%%%%%%%%%%%%%%%%%%%%%%%%%%%%%%%%%%

\begin{figure}[]
\includegraphics[clip,width=1.0\columnwidth]{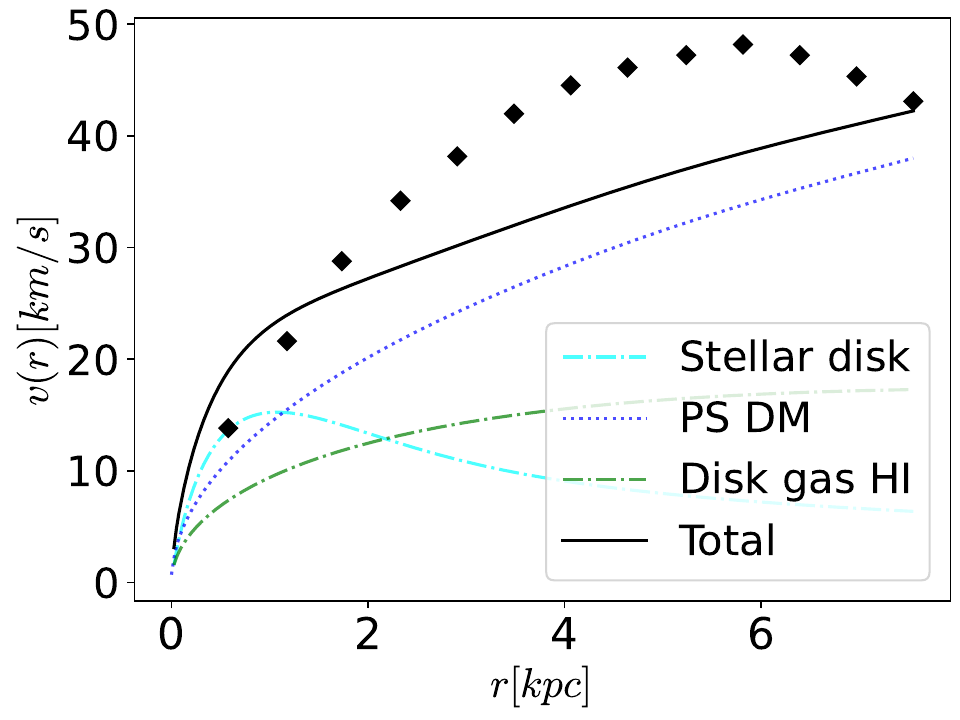}\\%
\includegraphics[clip,width=1.0\columnwidth]{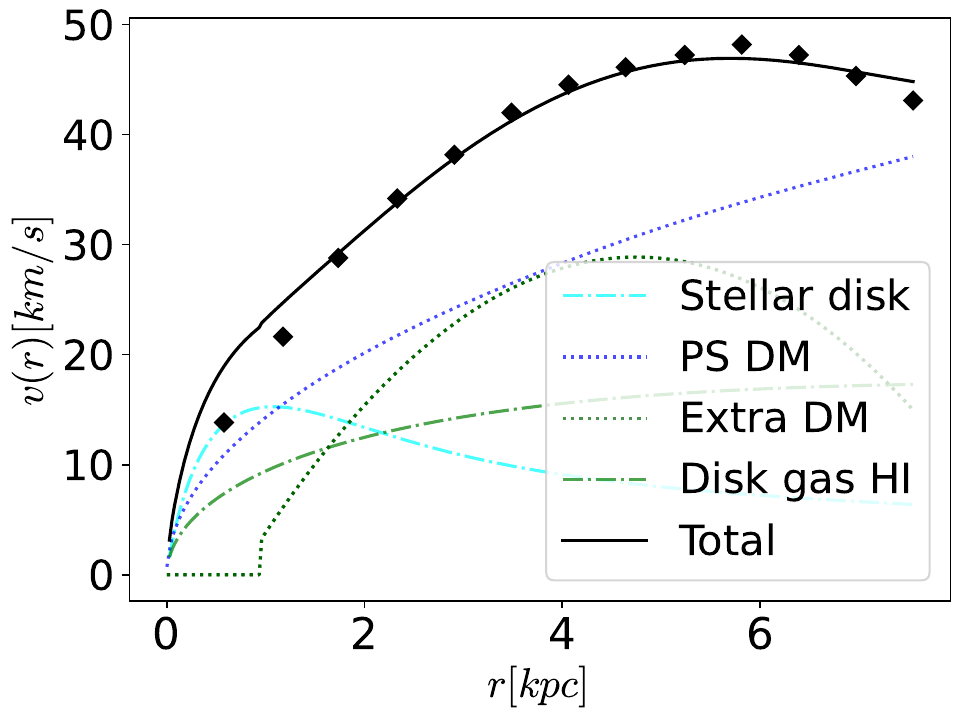}%
\caption{The upper plot shows the original fitting. The lower plot shows the fitting with a readjusting of the field frequency to lower values, for a better behavior for small radii. Concretely, we used the limit value $w/\mu=0.9999$. The lower plot shows the fitting with the Extra DM component. It is clear that taking into account this extra term with an unknown origin we recover the observed RC.}
\label{DDO154bis}
\end{figure}

A second halo component could enhance the fit near the galaxy’s center. Conversely, smaller and slower galaxies exhibit a different pattern. While our model aligns closely with the outer regions represented in the latter half of the data, it slightly overshoots in the inner halo regions. It appears that these smaller galaxies might benefit from an additional dark matter contribution, an Extra Dark Matter (EDM) that bolsters the outer regions, allowing for an adjustment of the EP component to better fit the central areas. This proposed approach is illustrated in  \cref{DDO154bis}.

In this approach, we present two options. First, we fit the model using the EP component, adjusted with a unique set of parameters that enable additional components to represent the data better. Then, we apply a full fitting that incorporates an extra DM component. We observe a recurring pattern by applying this method to some of our least accurately fitted galaxies, which previously included only one DM component. This pattern is described by a simple quadratic function that governs the behavior of the extra dark matter. The coefficients for this fitting function are listed in \cref{tabla2}.
\begin{equation}
    v_{EDM}=A_2 r^2+A_1 r+A_0.
\end{equation}
Just by adding the previous component to the total RC model, 
\begin{equation}
    v_T^2(r)=v_d^2(r)+v_g^2(r)+v_b^2(r)+v_P^2(r)+ v_{EDM}^2(r)
\end{equation}
 we get the corrected curves for small galaxies.
\begin{table}[ht]
\centering
\begin{tabular}{|c|c|c|c|}
\hline
\textbf{Galaxy} & \textbf{$A_2$} & \textbf{$A_1$} & \textbf{$A_0$} \\ \hline
\textit{NGC 1560} & -1.26936822 & 19.11489979 & -10.06725384 \\ \hline
\textit{DDO 170} & -0.82969423 & 13.91600334 & -11.16791058 \\ \hline
\textit{DDO 154} & -1.77881105 & 16.90364123 & -11.24755661 \\ \hline
\textit{NGC 3109} & -0.70249748 & 13.36677766 & -9.85825216 \\ \hline
\end{tabular}
\caption{Coefficients for the Extra Dark Matter fitting perform over the mentioned galaxies. As can be read from the values, for each of the orders there is kind of central value. }
\label{tabla2}
\end{table}
We can also read from \cref{tabla2} that all the coefficients for the fitting are very close. All the values are about $A_2\sim 1$, $A_1\sim 15$ and $A_0\sim -10$, meaning that the previously missing component is, in fact, following approximately the same behavior in all cases, showing some universality, that is, a good sign for our treatment. We can derive the mass profile and density generating this unknown DM component. With the parametrization for this curve component, we can do just the inverse calculation than in previous cases, 
\begin{equation}
\begin{split}
    &\sqrt{\frac{G M_{EDM}(r)}{r}}=A_2 r^2+A_1 r+A_0\\
    &\rightarrow M_{EDM}(r)=\frac{r}{G}(A_2 r^2+A_1 r+A_0)^2
    \end{split}
\end{equation}
With an associated density function obtained through the relation $\rho_{EDM}(r)=\frac{1}{2\pi r^2 }\frac{dM(r)}{dr}$
\begin{equation}
    \rho_{EDM}(r)=\frac{(r(A_2r+A_1)+A_0)(r(5A_2r+3A_1))+A_0} {2\pi r^2 G} .
\end{equation}
We can integrate this density and obtain the contribution of this component in the following way,
\begin{equation}
    M_{EDM}=\int_0^{r_{max}}2\pi r^2 \rho_{EDM}(r)dr .
\end{equation}
Through the previous integral we obtain for $M_{EDM}\sim 19.6\cdot 10^{11}M_{\odot}$ for \textit{NGC 1560}, $M_{EDM}\sim 17.7\cdot 10^{11}M_{\odot}$ for \textit{DDO 170},  $M_{EDM}\sim 1.9\cdot 10^{11}M_{\odot}$ for \textit{DDO 154} and   $M_{EDM}\sim 25.5\cdot 10^{11}M_{\odot}$ for \textit{NGC 3109}. All the obtained values are in the right order of magnitude for the DM halo contribution $\sim 10^{10}-10^{12} M_{\odot}$ for dwarf galaxies dominated by DM, which is a good signal.

To complete the test, we calculate the mass associated with the EP component. Initially, we reference \cref{masss} and find that for $w/\mu=0.9999$, the value is $M_A\mu=0.02363$. Assuming $\mu\sim 25\cdot 10^{-26}\mathrm{eV}$, this leads to a total mass for the EP system of approximately $M_A\sim 100\cdot 10^{11}\mathrm{M_{\odot}}$. This value is close to the values of $M_{EDM}$, slightly exceeding them, and falls within the observed range of total dark matter, which is consistent with expectations given that the EP contribution typically exceeds that of EDM.

A limitation of this EDM counterpart is that no observables can be linked to our fit, leaving the origin of this potential contribution uncertain. It might be generated by a static scalar boson field or a different type of particle associated with dark matter.

The requirement of an additional DM component for a precise description of dwarf galaxies might be seen as a drawback of our analysis. Here we want to argue that this is not a problem of our analysis in particular. Indeed, certain special galaxies, like dwarf spheroidals or ultra-diffuse galaxies, are well-known to require multiple dark matter components \cite{Burkert:1995yz,Bullock:2017xww} for a reasonable description. These galaxies often display unusual core-like dark matter density profiles or atypical rotational dynamics that are challenging to explain with a single dark matter species. 

For our concrete case, where one DM component is provided by vectorial bosonic matter,
a semi-quantitative argument can be inferred from the relation between the compactness and the quadrupolar moment for the studied scalar and vectorial models. Compactness simply is the ratio between mass and object radius. In the bosonic framework, we have to take the radius which encloses the $99\%$ of the total mass, and the corresponding mass,
\begin{equation}
    C=\frac{M_{99}}{R_{99}}.
\end{equation}
The quadrupolar moment, on the other hand, describes the amount of deformation under rotation. This magnitude tells us how the body spreads out or keeps its matter cohesive in space. The calculation of the quadrupolar moment is not trivial, and we refer to \cite{Adam:2024zqr} for a detailed study of this multipolar quantity.

\begin{figure*}[]
%\vspace*{-1cm}
\centering
\hspace*{-0.7cm}\includegraphics[width=0.50\textwidth]{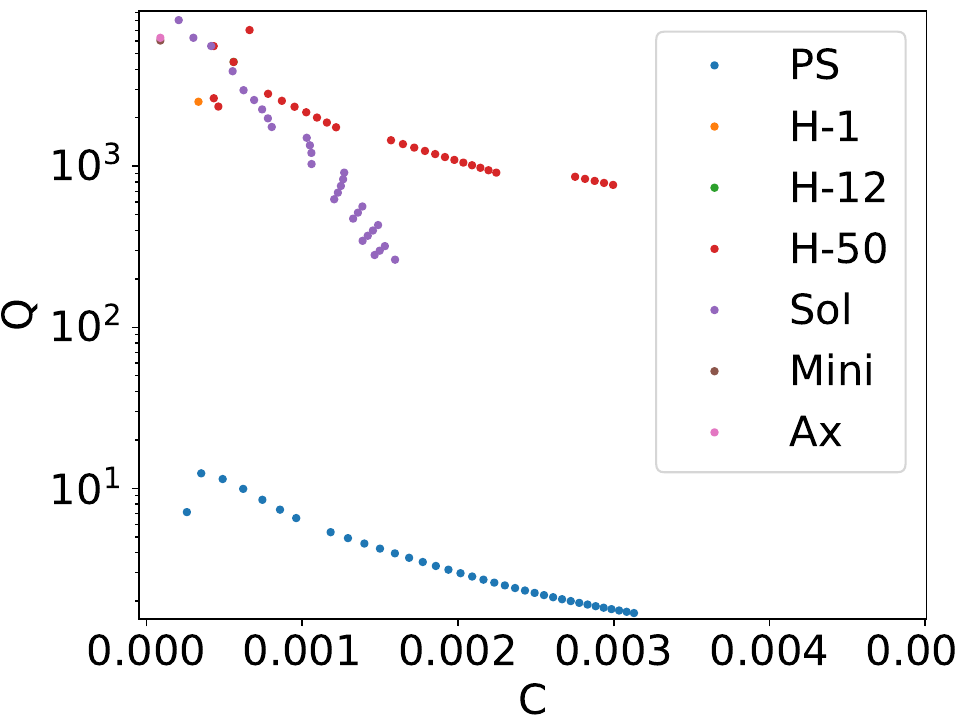}
\caption{Compactness vs quadrupolar deformation for the vector bosonic DM (PS) and for different scalar field DM models.}
\label{comparativa}
\end{figure*}

In \cref{comparativa}, we show the compactness of the considered models against the quadrupolar moment. We can see how the vectorial matter shows a significantly lower quadrupolar momentum for comparable compactness. This fact can be interpreted as follows: For comparable compactness, the vectorial matter tends to be more localized and robust against deformations. It is more cohesive and, as a result, occupies only the galactic space without spreading out of the galaxy. For galaxy species with small sizes and high compactness, we can see in \cref{comparativa} how the quadrupolar momentum for the vectorial case decays to lower values. This means that the shape of the galactic halo tends to be more spherical and compact, forming a kind of internal vectorial DM halo, which could sustain a second component of less compact and more deformable DM around it, which in the plot is represented by any of the scalar field DM models, for concreteness.

The same observation also serves to solve a second puzzle related to the best-fit values for the vector boson mass. Indeed, we found boson masses about $\mu \sim 1-3\cdot 10^{-25}\mathrm{eV}$. This should be contrasted with the values found from cosmological considerations for scalar DM in intergalactic space. In the framework of an ultra-light axion (ULA) DM model, e.g., Lyman-alpha forest data imply a lower bound of $\mu_{\rm ULA} \ge 10^{-22}\mathrm{eV}$ for the (scalar) axion mass, \cite{Rogers:2020ltq}. The cohesive character of the vectorial DM implies that it does not leak into the intergalactic space and, therefore, is irrelevant for the cosmological implications of intergalactic DM, which must be provided by other DM components.

%%%%%%%%%%%%%%%%%%%%%%%%%%%%%%%%%%%%%%%%%%%%%%%%%%%%%%%%%%%%%%
\section{Conclusions}
\label{conclusions}
%%%%%%%%%%%%%%%%%%%%%%%%%%%%%%%%%%%%%%%%%%%%%%%%%%%%%%%%%%%%%%
We studied the possibility to model dark matter galaxy halos using rotating scalar and vectorial bosonic fields coupled to gravity. We treated these DM candidates in a fully relativistic framework but in the non-relativistic regime $w/\mu\rightarrow 1$.

Contrary to the static case \cite{Schunck:1998nq,Urena-Lopez:2010zva,Chen:2020cef,Annulli:2020ilw}, the EKG system is not able to fit RC within the rotating scenario, as shown in \cref{curvesmodelsbs}. The reason is related to the excessively high rotation velocities produced by the rotating bosonic scalar matter in an axisymmetric space-time. This result obstructs the opportunity to proceed with this type of models, which has been successfully employed in other regimes and scales, such as BH or NS mimickers and intermediate mass astrophysical scales. This negative result could be related to the matter distribution for spinning rotating fields. As mentioned in the manuscript, the toroidal matter distribution makes the system more inert, and the matter around it rotates faster. Also, this toroidal distribution does not match perfectly with the observed galactic halos, which tend to be spherical or spheroidal.

In contrast with the previous case, the spinning vectorial bosonic fields are good candidates for fitting the galaxies we studied \cref{curvespsmodel}. The lower rotation curves found for the models proposed open the new possibility to parametrize DM with vectorial bosons. For various values of $w$, especially for those near the non-relativistic extreme, we found various curves with accurate shapes and maximal velocities \cref{procasmodelscurvessolos}. Through the role of the boson mass $\mu$, the scales can be adjusted to the specific galaxies under treatment, and together with the luminous matter components, the observed RC can be fitted.

In \cref{models}, we present the luminous models selected for our analysis, which are widely referenced in the literature. Utilizing the observational parameters detailed in \cref{table1}, we characterize the disk components of gas, stellar matter, and the bulge when relevant. The DM halo is incorporated into the total velocity using the EP system. Initially, the scale for each galaxy is determined by the boson mass. Subsequently, we select the frequency model which most accurately aligns with the chosen case.

The fittings we show in \cref{results1},  where the DM haloes are parametrized with EP models, are all at the right scales, with the correct maximal velocities and good accuracy. We notice, however, that some RCs are better fitted than others. An interesting result is that the ten studied galaxies can, in general, be split into two kinds: big spiral galaxies, which are impressively well adjusted by the EP halo alone, and small dwarf type galaxies with well-fitted distant regions and near halo cores, fitted with a certain excess. Our approach works without the need of any other treatment for the big galaxies, making the EP dark matter halos alone good candidates. What is more, similar sizes between galaxies imply the same range for the mass of the boson, and for this RC, we also have similar field frequencies, having an averaged $\mu\sim 9\cdot 10^{-26}\mathrm{eV}$ and $w/\mu\sim 0.9997$. This tendency to non-dispersed values for the parameters is also a good property of the model, making it stronger.

For the smaller, DM-dominated dwarf galaxies included in our study, along with other small galaxies, the EP fitting proved less accurate despite being within the correct size regime. We observed that if the outer part of the fit was appropriate, the inner part was excessively modeled. Consequently, we considered exploring a different approach, as detailed in \cref{results2}, where we adjusted the EP frequencies to better fit the inner regions, even if it meant an under-adjusting of the outer areas. This approach suggested the absence of a second DM component, which would significantly impact the distant regions. Through this new methodology, we subtracted the incomplete $v_T$ from the actual data and found a missing rotational curve component that followed a quadratic law. Upon examining the coefficients for the studied cases, we found them very close, indicating that similar underlying physics was responsible for the discrepancies. This quadratic velocity law enabled us to deduce a mass distribution and density for the unidentified DM component. Employing this traditional calculation method, we restored the alignment between the observed data and our model, now incorporating two complementary DM components: one from the EP system and another from an unknown source.

Interestingly, when calculating the total mass derived from the quadratic rotation law for these small galaxies, we found values consistent with the observational masses for these galaxies.

Our research indicates that vectorial bosonic fields, operating within the framework of Einstein gravity and governed by the Proca field equation, are effective at fully fitting the DM haloes of large spiral galaxies and partially fitting those of smaller and dwarf galaxies. Furthermore, we propose the presence of a secondary DM source characterized by a specific matter distribution. Integrating this source with the EP model could greatly improve the accuracy of our models for DM-dominated dwarf and small galaxies. This dual-source approach points to a more detailed and nuanced representation of the DM composition in these galaxies, suggesting a broader scope for understanding their dynamics and structure.

Various improvements can be made based on the findings and methodologies discussed in this papaer. First, the luminous models we used can be improved. While widely employed, they are relatively simplistic and not the most accurate ones. Conducting a purely non-relativistic analysis of the EP system could provide insights into different vectorial bosonic models. This would allow for the incorporation of additional self-interacting potentials within the Newtonian framework, enabling the exploration of alternative properties, shapes, and velocities for the curves. Furthermore, considering non-spinning EP models could also enrich our understanding.

Experimental measurements beyond those used in our study could provide additional information on the DM profiles and allow to examine the potential relationships between them, the EP system, and the mass distribution described by the quadratic law. These advancements could significantly deepen our comprehension of dark matter dynamics and its interaction with other astrophysical components.

We anticipate that our analysis will prove beneficial both theoretically and empirically. Theoretically, it should enhance our understanding of dark matter and the dynamics of galaxies. Empirically, it could facilitate the identification of the non-luminous components that are evident in the rotation curves but absent from the electromagnetic spectra. This dual approach should help bridge the gap between observed galactic behaviors and the theoretical models that attempt to explain them.

%\paragraph*{Summary.---}

\begin{acknowledgements}
 JCM thanks E.Radu and Etevaldo S.C. Filho for their crucial help with the FIDISOL/CADSOL package and also thanks M.Huidobro and A.G. Martín-Caro for further useful comments.  
Further, the authors acknowledge the Xunta de Galicia (Grant No. INCITE09.296.035PR and Centro singular de investigación de Galicia accreditation 2019-2022), the Spanish Consolider-Ingenio 2010 Programme CPAN (CSD2007-00042), and the European Union ERDF.
AW is supported by the Polish National Science Centre,
grant NCN 2020/39/B/ST2/01553.
.  JCM thank the Xunta de Galicia (Consellería de Cultura, Educación y Universidad) for funding their predoctoral activity through \emph{Programa de ayudas a la etapa predoctoral} 2021. JCM thanks the IGNITE program of IGFAE for financial support.
\end{acknowledgements}

%\bibliography{biblio}% Produces the bibliography via BibTeX.

%\newpage
%\vspace*{25cm}
%\newpage
%%%%%%%%%%%%%%%%%%%%%%%%%%%%%%%%%%%%%%%%%%%%%%%%%%%%%%%%%
\appendix
\section{}
\label{appendixA}

{\bf{Derivation of the curves:}}
\label{Supmat2}
For the derivation of the orbits, we follow \cite{Gourgoulhon:2010ju,Meliani:2015zta}.
Let us consider a particle $\mathcal{P}$ with mass $m<<M$ orbiting our bosonic matter, where $M$ is the system mass. With $\Vec{v}$ the four-velocity, its components are $\Vec{v}=\left(\frac{dt}{d\tau},\frac{dr}{d\tau},\frac{d\theta}{d\tau},\frac{d\psi}{d\tau}\right)$, where $\tau$ is $\mathcal{P}$ proper time. We have $\Vec{v}\cdot\Vec{v}=-1$. The four-momentum is $\Vec{p}=m\Vec{v}$ and $\Vec{p}\cdot\Vec{p}=-m^2$.
Assuming that $\mathcal{P}$ only suffers the gravity of $M$, it will follow timelike geodesics and $\nabla_{\Vec{p}}\Vec{p}=0$. As the spacetime is axially symmetric, we have the following conserved quantities associated with the Killing vectors of the problem ($\Vec{\xi}$ is the stationary symmetry associated and $\Vec{\chi}$ the axial one.):
\begin{equation}
    E := -\Vec{\xi}\cdot\Vec{p}=-p_t=cte,\hspace{0.2cm} L := \Vec{\chi}\cdot\Vec{p}=p_{\phi}=cte.
\end{equation}
The energy for a Zero Angular Momentum Observer is 
\begin{equation}
 E_{ZAMO}=e^{-\nu}\left(E-\frac{W}{r}L\right)  
 \label{zamoenergy}
\end{equation}
where we have used directly expression $(4.72)$ from \cite{Gourgoulhon:2010ju}.
Let us see how to derive the circular orbits.
In the equatorial plane, $\theta=\frac{\pi}{2}$ and $p^{\theta}=m\frac{d\theta}{d\tau}=0$. Through the metric,
\begin{equation}
    p_{\theta}=g_{\theta\mu}p^{\mu}=g_{\theta\theta}p^{\theta}=e^{2\alpha}r^2p^{\theta}=0.
\end{equation}

We have that $p_t,p_{\phi}$ and $p_{\theta}$ are constants. The remaining component, $p_r$ is obtained as,
\begin{equation}
        g^{\sigma\rho}p_{\sigma}p_{\rho}=-m^2
\end{equation}
and
\begin{equation}
    \begin{split}
    &-e^{-2\nu}p_t^2-2\frac{W}{r}e^{-2\nu}p_tp_{\phi}+e^{-2\alpha}p_r^2+\frac{e^{-2\alpha}}{r^2}p_{\theta}^2+\\
    &\left(\frac{e^{-2\beta}\csc^2\theta-e^{-2\nu}W^2}{r^2}\right)p_{\phi}^2=-m^2.
    \end{split}
\end{equation}
Using $\theta=\frac{\pi}{2},p_t=-E,p_{\theta}=0$ and $p_{\phi}=L$, we get,
\begin{equation}
  p_r^2e^{-2\alpha}=\left(E-\frac{W}{r}L\right)^2e^{-2\nu}-m^2-\frac{e^{-2\beta}}{r^2}L^2 
\end{equation}

having also
\begin{equation}
    p_r=g_{r\mu}p^{\mu}=e^{2\alpha}p^r=e^{2\alpha}m\frac{dr}{d\tau}
\end{equation}
We can join the two above expressions in the following,
\begin{equation}
\begin{split}
    &\frac{1}{2}\left(\frac{dr}{d\tau}\right)^2=\mathcal{V}(r,\bar{E},\bar{L})=\\
    &\frac{e^{-2\alpha}}{2}\left[\left(\bar{E}-\frac{W}{r}\bar{L}\right)^2e^{-2\nu}-1-\frac{e^{-2\beta}}{r^2}\bar{L}^2\right].
    \end{split}
    \label{effective}
\end{equation}
with  $\bar{E}=\frac{E}{m}$ and $\bar{L}=\frac{L}{m}$. The role of $\mathcal{V}(r,\bar{E},\bar{L})$ is to be the effective potential.
For obtaining the circular orbits, we have to fix $r=cte$ and have a minimum of the potential, so the next conditions arise,
\begin{equation}
    \frac{dr}{d\tau}=0,\hspace{0.3cm}\frac{dr^2}{d\tau^2}=0.
\end{equation}
This is equivalent, through \cref{effective}, to $\mathcal{V}(r,\bar{E},\bar{L})=0$ and $\frac{\partial \mathcal{V}}{\partial r}=0$ and we finish with
\begin{equation}
    \begin{split}
    &\frac{e^{-2\alpha}}{2}\left[\left(\bar{E}-\frac{W}{r}\bar{L}\right)^2e^{-2\nu}-1-\frac{e^{-2\beta}}{r^2}\bar{L}^2\right]=0,\\
    &
    \frac{\partial}{\partial r} \left[\left(\bar{E}-\frac{W}{r}\bar{L}\right)^2e^{-2\nu}-1-\frac{e^{-2\beta}}{r^2}\bar{L}^2\right]=0    
    \end{split}
\end{equation}
The last equation can be easily done; let's explicitly obtain the derivatives. Calling $A=\left(\bar{E}-\frac{W}{r}\bar{L}\right)^2e^{-2\nu}$ and $B=\frac{e^{-2\beta}}{r^2}\bar{L}^2$ we have:
\begin{equation}
\begin{split}
    &\frac{\partial A}{\partial r}=e^{-2\nu}\frac{\partial }{\partial r}\left(\bar{E}-\frac{W}{r}\bar{L}\right)^2+\left(\bar{E}-\frac{W}{r}\bar{L}\right)^2\frac{\partial }{\partial r}e^{-2\nu}=\\
    &=-2\bar{L}e^{-2\nu}\left(\bar{E}-\frac{W}{r}\bar{L}\right)\frac{\partial}{\partial r}\left(\frac{W}{r}\right)\\&
    -2e^{-2\nu}\left(\bar{E}-\frac{W}{r}\bar{L}\right)^2\frac{\partial \nu}{\partial r}=\\
    &
    =-2e^{-2\nu}\left(\bar{E}-\frac{W}{r}\bar{L}\right)\left[\left(\bar{E}-\frac{W}{r}\bar{L}\right)\frac{\partial \nu}{\partial r}+\bar{L}\frac{\partial}{\partial r}\left(\frac{W}{r}\right)\right]
    \end{split}
\end{equation}
The other term,
\begin{equation}
    \frac{\partial B}{\partial r}=\bar{L}^2\left[\frac{-2r^2e^{-2\beta}\frac{\partial \beta}{\partial r}-2re^{-2\beta}}{r^4}\right]=-2\frac{\bar{L}^2e^{-2\beta}}{r^2}\left[\frac{\partial \beta}{\partial r}+\frac{1}{r}\right].
\end{equation}
The addition of the two previous terms and by erasing the common factors, as the full expression is equated to zero, we arrive at,
\begin{equation}
\begin{split}
    &e^{-2\nu}\left(\bar{E}-\frac{W}{r}\bar{L}\right)\left[\left(\bar{E}-\frac{W}{r}\bar{L}\right)\frac{\partial \nu}{\partial r}+\bar{L}\frac{\partial}{\partial r}\left(\frac{W}{r}\right)\right]-\\&
    \frac{\bar{L}^2e^{-2\beta}}{r^2}\left[\frac{\partial \beta}{\partial r}+\frac{1}{r}\right]=0.
    \end{split}
    \label{derivative}
\end{equation}
The particle angular velocity seen by a distant observer is,
\begin{equation}
    \Omega_{\mathcal{P}}=\frac{v^{\phi}}{v^t},
\end{equation}
We will use the Killing vectors and some $3+1$ tools to obtain the modulus of the particle $3-velocity$. First of all, under the circularity conditions (The Generalized Papapetrou theorem), we can write the particle four-velocity as a combination of the two Killings through the angular velocity and the velocity time component,
\begin{equation}
    \Vec{v}=v^t\left(\Vec{\xi}+\Omega_{\mathcal{P}}\Vec{\chi}\right)
\end{equation}
Now, let us link our expression with the $3+1$ formalism. The ZAMO observers are Eulerian observers in the context of the rotating star, whose worldlines are everywhere orthogonal to the spatial hypersurfaces for a given slicing \cite{Gourgoulhon:2007ue}. So, their four velocities are unit timelike vectors.
Now let us reexpress our particle velocity in such decomposition, formed by a ZAMO four-velocity $\Vec{n}$ and a particle space three-velocity $\Vec{V}$,
\begin{equation}
    \Vec{v}=v^te^{\nu}(\Vec{n}+\Vec{V}).
\end{equation}
The previous, together with the spatial killing decomposition in the $3+1$ way in terms of the hypersurface orthogonal $\Vec{n}$ and tangent part, $\Vec{\xi}=e^{\nu}\Vec{n}-\frac{W}{r}\Vec{\chi}$ allows us to obtain the expression for the three-velocity,
\begin{equation}
\begin{split}
    &\Vec{\xi}+\Omega_{\mathcal{P}}\Vec{\chi}=e^{\nu}(\Vec{n}+\Vec{V})\rightarrow e^{\nu}\Vec{n}-\frac{W}{r}\Vec{\chi}+\Omega_{\mathcal{P}}\Vec{\chi}=e^{\nu}\Vec{n}+e^{\nu}\Vec{V}\\&
    \rightarrow\Vec{V}=e^{-\nu}\left(\Omega_{\mathcal{P}}-\frac{W}{r}\right)\Vec{\chi}
    \end{split}
\end{equation}
Since $\Vec{\chi}\cdot\Vec{\chi}=e^{2\beta}r^2\sin^2{\theta}$, and $\theta=\pi/2$, we obtain the modulus for the three-velocity,
\begin{equation}
    V=re^{\beta-\nu}\left(\Omega_{\mathcal{P}}-\frac{W}{r}\right).
\end{equation}

Now let us see the expression for $\bar{E}=\frac{E}{m}=-\frac{p_t}{m}=m\frac{-v_t}{m}=-v_t$;
\begin{equation}
\begin{split}
    &\bar{E}=-v_t=-g_{t\mu}v^{\mu}=-g_{tt}v^t-g_{t\phi}v^{\phi}=v^t(-g_{tt}-\Omega_{\mathcal{P}}g_{t\phi})=\\
    &
    v^t\left(e^{2\nu}+e^{2\beta}Wr\left[\Omega_{\mathcal{P}}-\frac{W}{r}\right]\right)=e^{\nu}v^t\left(e^{\nu}+e^{\beta}VW\right)\rightarrow\\
    &
    \bar{E}=e^{\nu}v^t\left(e^{\nu}+e^{\beta}VW\right)=\Gamma_{\mathcal{P}}\left(e^{\nu}+e^{\beta}VW\right)
    \end{split}
\end{equation}
Where we have identified the term $e^{\nu}v^t=\Gamma_{\mathcal{P}}$ that is the Lorentz factor for the ZAMO.
In the same manner, we have,
\begin{equation}
    \begin{split}
        &\bar{L}=v_{\phi}=g_{\phi\mu}v^{\mu}=g_{\phi\phi}v^{\phi}+g_{\phi t}v^t=\\
        &
        =v^t\Omega_{\mathcal{P}}g_{\phi\phi}+g_{\phi t}v^t=v^t\left[e^{2\beta}r^2\Omega_{\mathcal{P}}-e^{2\beta}rW\right]=\\
        &
        v^te^{2\beta}r^2\left[\Omega_{\mathcal{P}}-\frac{W}{r}\right]=\\
        &
        =v^te^{\nu}e^{\beta}rV\rightarrow\bar{L}=\Gamma_{\mathcal{P}}e^{\beta}rV.
    \end{split}
    \label{L}
\end{equation}

If we multiply $W/r$ by $\bar{L}$ and we add $\bar{E}$,
\begin{equation}
    \bar{E}+\frac{W}{r}\bar{L}=\Gamma_{\mathcal{P}}e^{\nu}
    \label{EL}
\end{equation}
If we retake \ref{zamoenergy} and use the above, we get,
\begin{equation}
    E_{ZAMO}=m\Gamma_{\mathcal{P}}.
\end{equation}
Now we introduce \ref{L},\ref{EL} in equation $\mathcal{V}(r,\bar{E},\bar{L})=0$ and what we easily obtain is,
\begin{equation}
    V^2\Gamma_{\mathcal{P}}^2+1-\Gamma_{\mathcal{P}}^2=0\rightarrow\Gamma_{\mathcal{P}}=\frac{1}{\sqrt{1-V^2}},
\end{equation}
as expected.
And by doing the same substitutions in equation \ref{derivative}, we get,
\begin{equation}
  \begin{split}
    &\Gamma_{\mathcal{P}}\left[\Gamma_{\mathcal{P}}\frac{\partial \nu}{\partial r}+\Gamma_{\mathcal{P}}e^{\beta-\nu}rV\frac{\partial}{\partial r}\left(\frac{W}{r}\right)\right]-\Gamma_{\mathcal{P}}^2V^2\left(\frac{\partial \beta}{\partial r}+\frac{1}{r}\right)  =0\\
    &\rightarrow V^2\left(\frac{\partial \beta}{\partial r}+\frac{1}{r}\right) - Ve^{\beta-\nu}r\frac{\partial}{\partial r}\left(\frac{W}{r}\right)-\frac{\partial \nu}{\partial r}=0
  \end{split}  
\end{equation}
We finally arrive at the velocity function, that is:
\begin{equation}
\begin{split}
    &V\pm=\frac{e^{\beta-\nu }r\frac{\partial}{\partial r}\left(\frac{W}{r}\right)\pm\sqrt{D}}{2\left(\frac{\partial \beta}{\partial r}+\frac{1}{r}\right)},\\
    &D=e^{2\beta-2\nu }r^2\left(\frac{\partial}{\partial r}\left(\frac{W}{r}\right)\right)^2+4\left(\frac{\partial \beta}{\partial r}+\frac{1}{r}\right)\frac{\partial \nu}{\partial r}.
    \end{split}
    \label{velocitystars}
\end{equation}

 \vspace{6cm}
 
\bibliography{biblio}

\end{document}